\documentclass[preprint,12pt]{elsarticle}

\usepackage{amssymb}
\usepackage{amsthm}
\usepackage{amsmath}
\usepackage{pbox}
\usepackage{booktabs}
\usepackage{color}
\usepackage{nicefrac}

\usepackage{lineno}
\usepackage{float}
\journal{Journal of Biological Physics}
\begin{document}

\begin{frontmatter}



\title{Adaptive stimulus design for dynamic recurrent neural network models}


\author[jhmi1,atlm]{R.Ozgur DORUK\corref{cor1}\fnref{grant1}}
\cortext[cor1]{Corresponding Author}
\fntext[label2]{This work is partially supported by Turkish Scientific and Technological Research Council (T\"{U}B\.{I}TAK) 2219 Research Program.}
\ead{resat.doruk@atilim.edu.tr}
\author[jhmi2]{Kechen Zhang}
\ead{kzhang4@jhmi.edu}
\address[atlm]{Atilim University, Department of Electrical and Electronics Engineering, Kizilcasar Mahallesi, Incek, Golbasi, Ankara, 06836,TURKEY}
\address[jhmi1]{Department of Biomedical Engineering, Johns Hopkins School of Medicine, 720 Rutland Avenue, Ross 528, Baltimore , MD, 21205, USA}
\address[jhmi2]{Department of Biomedical Engineering, Johns Hopkins School of Medicine, 720 Rutland Avenue, Traylor 407, Baltimore , MD, 21205, USA}

\begin{abstract}
We present a theoretical application of an optimal experiment design (OED) methodology to the development of mathematical models to describe the stimulus-response relationship of sensory neurons. Although there are a few related studies in the computational neuroscience literature on this topic, most of them are either involving non-linear static maps or simple linear filters cascaded to a static non-linearity. Although the linear filters might be appropriate to demonstrate some aspects of neural processes, the high level of non-linearity in the nature of the stimulus-response data may render them inadequate. In addition, modelling by a static non-linear input - output map may mask important dynamical (time-dependent) features in the response data. Due to all those facts a non-linear continuous time dynamic recurrent neural network that models the excitatory and inhibitory membrane potential dynamics is preferred. The main goal of this research is to estimate the parametric details of this model from the available stimulus-response data. In order to design an efficient estimator an optimal experiment design scheme is proposed which computes a pre-shaped stimulus to maximize a certain measure of Fisher Information Matrix. This measure depends on the estimated values of the parameters in the current step and the optimal stimuli are used in a maximum likelihood estimation procedure to find an estimate of the network parameters. This process works as a loop until a reasonable convergence occurs. The response data is discontinuous as it is composed of the neural spiking instants which is assumed to obey the Poisson statistical distribution. Thus the likelihood functions depend on the Poisson statistics. The model considered in this research has universal approximation capability and thus can be used in the modelling of any non-linear processes. In order to validate the approach and evaluate its performance, a comparison with another approach on estimation based on randomly generated stimuli is also presented. 
\end{abstract}

\begin{keyword}
Optimal Design, Sensory Neurons, Recurrent Neural Network, Excitatory Neuron, Inhibitory Neuron, Maximum Likelihood Estimation

\end{keyword}

\end{frontmatter}


\section{Introduction}
Optimal experiment design (OED) or shortly optimal design \citep{fedorov2013optimal} is a sub-field of optimal control theory which concentrates design of an optimal control law aiming at the maximization of the information content in the response of a dynamical system related to its parameters. The statistical advantage brought by information maximization helps the researchers to generate the best input to their target plant/system that can be used in a system identification experiment producing estimates with minimum variance \citep{telen2012approximate}. With the utilization of mathematical models in theoretical neuroscience research, the application of optimal experiment design in adaptive stimuli generation should be beneficial as it is expected to have better evaluations of the model specific parameters from the collected stimulus-response data. Though these benefits, the optimal experiment design have not found its place among theoretical or computational neuroscience research due to the nature of the models. As the stimulus-response relationship is naturally quite non-linear, computational complexity of the optimization algorithms utilized for an optimal experiment design will typically be very high and thus OED has not gained enough attraction during the past decades. However, thanks to the today's computational powers of new microprocessors, it will be much easier to talk about a real optimal experiment design in neuroscience research (\citep{benda2007response},\citep{newman2012closed}). In the past decades, some researchers had stimulated their models by Gaussian white noise stimuli \citep{paninski2003noise}, \citep{marmarelis1978analysis} and performed an estimation of input-output relationships of their model (\citep{chichilnisky2001simple} \citep{paninski2004maximum} and \citep{wu2006complete}). This algorithmically simpler approach is theoretically proven to be efficient in the estimation of models based on linear filters and their cascades. However, in \citep{dimattina2010modify}, it is suggested that white noise stimuli may not be successful as a stimuli in the parametric identification of non-linear response models due to high level of parameter confounding (refer to \citep{dimattina2013adaptive} for a detailed description of the confounding phenomenon in non-linear models). 

Concerning the applications of optimal experiment design to biological neural network models, there exist a limited amount of research. One such example is \citep{dimattina2011active} where a static non-linear input output mapping is utilized as a neural stimulus-response model. The optimal design of the stimuli is performed by the maximization of the D-Optimal metric of the Fisher Information Matrix (FIM) \citep{pukelsheim1993optimal} which reflects a minimization of the variance of the total parametric error of the model network. In the last research,the parameter estimation is based on the Maximum A Posteriori (MAP) Estimation methodology \citep{degroot1970optimal} which is linked to the Maximum Likelihood Estimation (ML) \citep{myung2003tutorial} approach. Two other successful mainly experimental work on applications of optimal experiment design to adaptive data collection are \citep{mseyaldekel} and \citep{phdtamw}. The experimental works successfully proven the efficiency of optimal designs for certain models in theoretical neuroscience. However, none of those studies explore fully dynamical non-linear models explicitly. Because of this deficiency, this research will concentrate on an application of the optimal experiment design to a fully dynamical non-linear model. The final goal is almost similar to that of \citep{dimattina2011active}.

The proposed model is a continuous time dynamical recurrent neural network (CTRNN) \citep{beer1995dynamics} in general and it also represents the excitatory and inhibitory behaviours \citep{ledoux2011dynamics} of the realistic biological neurons. Like in that of \citep{hodgkin1952quantitative} and its derivatives, the CTRNN describes the dynamics of the membrane potentials of the constituent neurons. However, the channel activation dynamics is not directly represented. Instead it constitutes, a more generic model which can be applied to a network having any number of neurons. The dynamic properties of the neuron membrane is represented by time constants and the synaptic excitation and inhibition are represented as network weights (scalar gains). Though not the same, a similar excitatory-inhibitory structure is utilized in numerous studies such as \citep{hancock1997modeling,hancock1999wideband,de2008linking}. As there isn't sufficient amount of research on the application of OED to dynamical neural network models, it will be convenient to start with a basic network model having two neurons representing the average of excitatory and inhibitory populations respectively. The final goal is to estimate the time constants and weight parameters. The optimal experiment design will be performed by maximizing a certain metric of the FIM. The FIM is a function of stimulus input and network parameters. As the true network parameters are not known in the actual problem, the Information Matrix should depend on the estimated values of the parameters in the current step. An optimization on a time dependent variable like stimulus will not be easy and often its parametrization is required. In auditory neuroscience point of view, that can be done by representing the stimuli by a sum of phased cosine elements. If periodic stimulation is allowed, these can be formed as harmonics based on a base stimulation frequency. The optimally designed stimulus will be the driving force of a joint maximum likelihood estimation (JMLE) process which involves all the recorded response data. Unfortunately, the recorded response data will not be continuous. The reason for this is that, in vivo measurements of the membrane potentials are often very difficult and dangerous as the direct manipulation with the neuron in vivo may trigger the death of a neuron. Thus, in the real experimental set-up, the peaks of the membrane potentials are collected as firing instants. As a result, one will only have a neural spike train with the exact neural spiking times (timings of the membrane potential peaks) but no other data. This outcome prevents one to apply traditional parameter estimation techniques such as minimum mean square estimation (MMSE) as it will require continuous firing rate (is based on the membrane potential) data. Researches like \citep{shadlen1994noise}, suggests that the neural spiking profile of sensory neurons obey the famous inhomogeneous Poisson distribution \citep{lewis1979simulation}. Under this assumption, the Fisher Information Matrix \citep{dimattina2011active} and Likelihood functions \citep{uteden2008pointprocesses,brown2002time} can be derived based on Poisson statistical distribution. The optimization of a certain measure of Fisher Information Matrix and the Likelihood can be performed by readily available packages such as MATLAB\textsuperscript{\circledR} Optimization Toolbox (like well known \emph{fmincon} algorithm).

There are certain challenges in this research. First of all, the limited availability of similar studies lead to the fact that this work is one of the first contributions on the applications of optimal experiment design to the dynamical neural network modelling. Secondly, we will most probably not be able to have a reasonable estimate just from a single spiking response data set as we do not have a continuous response data. This is also demonstrated in the related kernel density estimation research such as \citep{nawrot1999single,shimazaki2010kernel,shimazaki2007method,koyama2004histogram}. From these sources, one will easily note that repeated trials and superimposed spike sequences are required to obtain a meaningfully accurate firing rate information from the neural response data. In a real experiment environment, repeating the trials with the same stimulus profile will not be appropriate as the repeated responses of the same stimulus are found to be attenuated.
Because of this issue, a new stimulus should be designed each time based on the currently estimated parameters of the model and then it should be used in an updated estimation. These updated parameters are used in the next step to generate the new optimal stimulus. As a result one will have a new stimulus in each step and thus the risk of response attenuation is largely reduced. In a maximum likelihood estimation, the likelihood function will depend on the whole spiking data obtained throughout the experiment (or simulation). The parallel processing capabilities of MATLAB\textsuperscript{\circledR} (i.e. \emph{parfor}) on multiple processor/core computers will help in resolving of those issues.

\section{Models \& Methods}

\subsection{Continuous Time Recurrent Neural Networks} \label{sub:ctrnn-theory}
\begin{figure}[H]
\centering
\includegraphics[scale=0.7]{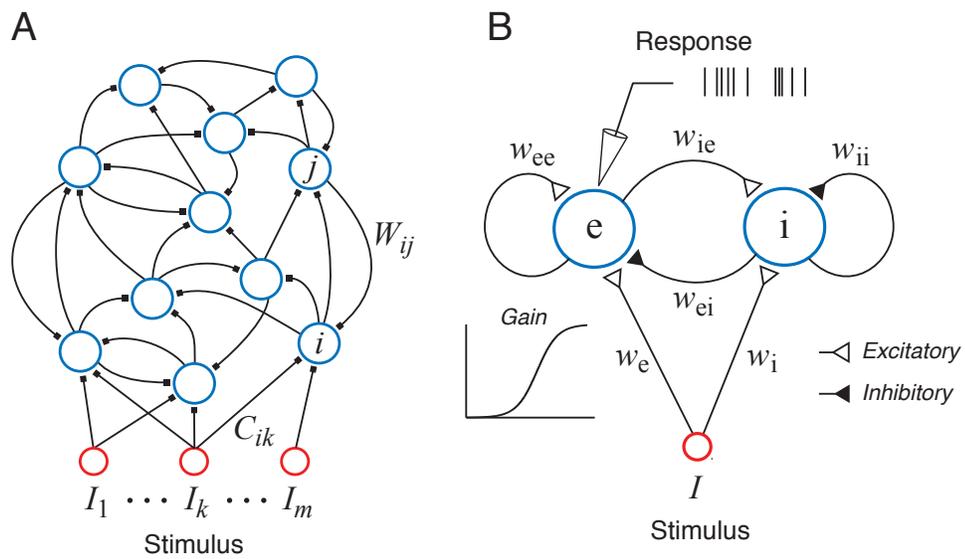}
\caption{(\textbf{A}) A generic recurrent neural network structure. The stimulus means external inputs to the network.
(\textbf{B}) A simple recurrent network with one excitatory unit and one inhibitory unit, with both units having nonlinear sigmoidal gain functions. Here each unit may represent a population of neurons.We assume that the recorded responses are inhomogeneous Poisson spike trains based on the continuous rate generated by the state of the excitatory unit.  }
\label{fig:generic-ctrnn}
\end{figure}

The continuous time recurrent neural networks have a similar structure to that of the discrete time counterparts that are often met in artificial intelligence studies. In \textbf{Figure \ref{fig:generic-ctrnn}}, one can see a general continuous time network that may have any number of neurons.

The mathematical representation of this generic model can be written as shown below \cite{beer1995dynamics}:
\begin{equation}
	 \tau_{i}\frac{dV_{i}}{dt}=-V_{i}+\sum_{j=1}^{n} W_{ij}g_{j}\left(V_{j}\right) +\sum_{k=1}^{m} C_{ik}I_{k}
	\label{eq:general-network-model}
\end{equation}
\noindent where ${\tau_k}$ is the time constant, ${V_k}$ is the membrane potential of the ${k^{th}}$ neuron, ${W_{kj}}$ is the synaptic connection weight between the ${k^{th}}$ and ${j^{th}}$ neurons ${C_{ki}}$ is the connection weight from ${i^{th}}$ input to the ${k^{th}}$ neuron and ${I_k}$ is the ${i^{th}}$ input. The term ${g_j\left(V_j\right)}$ is a membrane potential dependent function which acts as a variable gain on the synaptic inputs to from the ${j^{th}}$ neuron to the ${k^{th}}$ one. It can be shown by a logistic sigmoid function which can be shown as:
\begin{equation}
	 g_{j}\left(V_{j}\right)=\frac{\Gamma_j}{1+\exp\left( -a_{j}\left(V_{j}-h_j\right)\right) }\label{eq:sigmoid-general}
\end{equation} 
where $\Gamma_j$ is the maximum rate at which the ${j^{th}}$ neuron can fire, $h_j$ is a soft threshold parameter of the ${j^{th}}$ neuron and ${a_j}$ is a slope constant. This is the only source of non-linearity in \eqref{eq:general-network-model}. In addition it also models the activation-inactivation behaviour in more specific models of the neuron (like  \citep{hodgkin1952quantitative}). The work by \citep{miller2012mathematical} shows that \eqref{eq:sigmoid-general} gives a relationship between the firing rate ${r_j}$ and membrane potential ${V_j}$ of the ${j^{th}}$ neuron. In sensory nervous system, some of neurons have excitatory synaptic connections while some have inhibitory ones. This fact is reflected to the model in \eqref{eq:general-network-model} by assigning negative values to the weight parameters which are originating from neurons with inhibitory synaptic connections. In the introduction of this research, it is stated that it would be convenient to apply the theory to a basic network first of all due to the lack of related research and computational complexity. So a basic excitatory and inhibitory continuous time recurrent dynamical network can be written as shown in the following:
\begin{eqnarray}
	\tau_{e}\dot{V}_{e}&=&-V_{e}+w_{ee}g_{e}\left(V_{e}\right)-w_{ei}g_{i}\left(V_{i}\right)+w_{e}I
	\label{eq:2-exc-inh-network-a}
	\\
	\tau_{i}\dot{V_{i}}&=&-V_{i}++w_{ie}g_{e}\left(V_{e}\right)-w_{ii}g_{i}\left(V_{i}\right)+w_{i}I
	\label{eq:2-exc-inh-network-b}
\end{eqnarray} 
where the subscripts ${'e'}$ and ${'i'}$ stands for excitatory and inhibitory neurons respectively. Starting from now on, we will have a single stimulus and it will be represented by the term ${I}$ which will be generated by the optimal design algorithm. In addition in order to suit the model equations to the estimation theory formalism the time constant may be moved to the right hand side as shown below:

\begin{equation} 
	\frac{d}{dt}\left[\begin{array}{c}
	V_{e}\\
	V_{i}
		\end{array}\right]=\left[\begin{array}{cc}
		\beta_{e} & 0\\
		0 & \beta_{i}
		\end{array}\right]\left\{ -\left[\begin{array}{c}
		V_{e}\\
		V_{i}
		\end{array}\right]+\left[\begin{array}{cc}
		w_{ee} & -w_{ei}\\
		w_{ie} & -w_{ii}
		\end{array}\right]\left[\begin{array}{c}
		g_{e}\left(V_{e}\right)\\
		g_{i}\left(V_{i}\right)
		\end{array}\right]+\left[\begin{array}{c}
		w_{e}\\
		w_{i}
		\end{array}\right]I\right\} \label{eq:our-model-matrix-form}
\end{equation}
\textcolor{red}{where $ \beta_e $ and $ \beta_i $ are the reciprocals of the time constants $ \tau_e $ and $ \tau_i $. They are taken to the right for easier manipulations of the equations. }	
\noindent Note that this equation is written in matrix form to be conformed to the formal non-linear system forms. A descriptive illustration related to \eqref{eq:our-model-matrix-form} is presented in \textbf{Figure \ref{fig:generic-ctrnn}b}. It should also be noted that, in \eqref{eq:2-exc-inh-network-b} and \eqref{eq:our-model-matrix-form} the weights are all assumed as positive coefficients and they have signs in the equation. So negative signs indicate that originating neuron is inhibitory (tend to hyper-polarize the other neurons in the network).

\subsection{Inhomogeneous Poisson spike model}\label{Sub:poisson-theory} 

The theoretical response of the network in \eqref{eq:2-exc-inh-network-b} will be the firing rate of the excitatory neuron as ${r_e=g_e\left(V_e\right)}$. In the actual environment, the neural spiking due to the firing rate ${r_e\left(t\right)}$ is available instead. While introducing this research, it is stated that this spiking events conform to an inhomogeneous Poisson process which is defined below:
\begin{equation}
	\mbox{Prob}\left[N\left(t+\Delta t\right)-N\left(t\right)=k\right]=\frac{e^{-\lambda} \lambda^{k}}{k!}
	\label{eq:inhomogeneous-poisson}
\end{equation} 
where
\begin{equation}
	\lambda=\int_{t}^{t+\Delta t}r_e\left(\tau\right)d\tau
\end{equation}
is the mean number of spikes based on the firing rate $r_e(t)$ which varies with time,  and $N(\tau)$ indicates the cumulative total number of spikes up to time $\tau$, so that  $N\left(t+\Delta t\right)-N\left(t\right)$ is the number of spikes within the time interval ${\left[t,t+\Delta t\right)}$. 
In other words, the probability of having ${k}$ number of spikes in the interval ${\left(t,t+\Delta t\right)}$ is given by the Poisson distribution above. 

Consider a spike train  $(t_1,t_2,\ldots,t_K)$ in the time interval $(0, T)$ \textcolor{red}{(here $ 0\leq t_1 \leq t_2 \leq \ldots \leq t_K \leq T $ so $ t $ and $ \Delta t $ become $ t=0 $ and $ \Delta t=T $)}.  Here the spike train is described by a list of the time stamps for the $K$ spikes.  
The probability density function for a given spiking train $(t_1,t_2,\ldots,t_K)$ can be derived from the inhomogeneous Poisson process  \citep{uteden2008pointprocesses,brown2002time}. The result reads:
\begin{equation}
 	p\left(t_{1},t_{2},\ldots,t_{K}\right)
	=\exp\left(-\int_{0}^{T}r_e\left(t\right)dt\right)\prod_{k=1}^K r_e\left(t_k,{\bf x},\theta\right)
	\label{eq:lkl-enbrown}
\end{equation}     
This probability density describes how likely a particular spike train $(t_1,t_2,\ldots,t_K)$ is generated by the inhomogeneous Poisson process with the rate function $r_e\left(t,{\bf x},\theta\right)$. 
Of course, this rate function depends implicitly on the network parameters and the stimulus used. 
\subsection{Maximum Likelihood Methods and Parameter Estimation} \label{Sub:jmle-theory}

The network parameters to be estimated are listed below as a vector:
\begin{equation}
 \theta=\left[\theta_1, \ldots,\theta_8\right]
 =\left[\beta_e,\beta_i,w_e,w_i,w_{ee},w_{ei},w_{ie},w_{ii}\right]\label{eq:theta-ctrnn-param}
\end{equation}
which includes the time constants and all the connection weights in the E-I network.
Our maximum-likelihood estimation of the network parameters is based on the likelihood function given by \eqref{eq:lkl-enbrown}, 
which takes the individual spike timings into account. 
It is well known from estimation theory is that maximum likelihood estimation is asymptotically efficient, i.e., reaching the Cram\'er-Rao bound in the limit of large data size.

To extend the likelihood function in \eqref{eq:lkl-enbrown} to the situation where there are multiple spike trains elicited by multiple stimuli,
consider a sequence of $M$ stimuli. 
Suppose the $m$-th stimulus ($m=1,\ldots, M$) elicits a spike trains with a total of $K_m$ spikes in the time window $[0,T]$, and the spike timings are given by
 ${S}_m=\left(t_1^{(m)},t_2^{(m)},\ldots,t_{K_m}^{(m)}\right)$. 
 By \eqref{eq:lkl-enbrown}, the likelihood function for the spike train $S_m$ is
 \begin{equation}
 	p\left(S_m\mid\theta\right)
	=\exp\left(-\int_{0}^{T}r_e^{(m)}\!\left(t\right)dt\right)\prod_{k=1}^{K_m} r_e^{(m)}\!\left(t_{k}^{(m)}\right)
	\label{eq:pSm}
\end{equation}     
where $r_e^{(m)}$ is the firing rate in response to the $m$-th stimulus. Note that the rate function $r_e^{(m)}$  depends implicitly on the network parameters $\theta$ and on the stimulus parameters. The left-hand side of \eqref{eq:pSm} emphasizes the dependence on network parameters $\theta$, which is convenient for parameter estimation. 
The dependence on the stimulus parameters will be discussed in the next section.

We assume that the responses to different stimuli are independent, which is a reasonable assumption when the inter-stimulus intervals are sufficiently large. Under this assumption, the overall likelihood function for the collection of all $M$ spike trains can be written as
 \begin{equation}
	 L\left({S}_{1},S_{2},\ldots,S_{M} \mid \theta \right)
	 =\prod_{m=1}^{M}p\left(S_m \mid\theta\right)
	 \label{eq:joint-likelihood-product}
\end{equation}
By taking natural logarithm, we obtain the log likelihood function:
\begin{equation}
	l \left({S}_{1},S_{2},\ldots,S_{M} \mid \theta\right)
	=-\sum_{m=1}^{M} \int_{0}^{T}r_e^{(m)}\!\left(t\right)dt 
	+\sum_{m=1}^{M}\sum_{k=1}^{K_m}\ln{r_e^{(m)}}\!\left(t_{k}^{(m)}\right)
	\label{eq:complete-likelihood-compact}
\end{equation}
Maximum-likelihood estimation of the parameter set is given formally by
\begin{equation}
 	\hat\theta_{ML}=\arg\max_{\theta} l \left({S}_{1},S_{2},\ldots,S_{M} \mid \theta\right)
	\label{eq:mle-arg-max-log}
\end{equation}
Numerical issues related to this optimization problem will be discussed in \textbf{Sections \ref{sub:grad-computation}} and \textbf{\ref{sub:other-numerical}}. In addition, some discussion on the local maxima problems is provided in \textbf{Section \ref{sub:local-maxima-discussion}}. 
\subsection{Objective function of optimal design of stimuli}\label{Sub:oed-theory}

The optimal design method generates the stimuli by maximizing a utility function, or an objective function. 
The basic idea is that these stimuli are designed so as to elicit responses that are most informative about the network parameters.
In optimal design method, the utility function $U({\bf x},\theta)$ depends on the stimulus parameters $\bf x$, but typically also on the model parameters $\theta$.
An intuitive explanation of the dependence on the model parameter is best illustrated with an example. 
Suppose we want to estimate a Gaussian tuning curve model with unknown parameters although we may have some idea about the sensible ranges of these parameters. To estimate the height of the tuning curve accurately, we should place a probing stimulus around the likely location of the peak. To estimate the width, the probing stimulus should go to where the tuning curve is likely to have the steepest slope. For the baseline, we should go for the lowest response. This simple example illustrates two facts: first, optimal design depends on our knowledge of possible parameter values; second, the elicited responses in an optimally design experiment are expected to vary over a wide dynamic range as different parameters are estimated.

Once the utility function $U({\bf x},\theta)$ is chosen, the optimally designed stimulus may be written formally as:
\begin{equation}
	\hat{\bf x}=\arg\max_{\bf x} U({\bf x},\theta)\label{eq:oed-general-def}
\end{equation}
where the network parameters $\theta$ can be obtained by maximum-likelihood estimation from the existing spike data as described in the preceding section. 
Here the stimulus is specified by vector $\bf x$, which is a set of parameters rather than the actual stimulus itself.
Direct computation of the actual time-varying stimulus is not easy because no closed analytical form of the objective function is available and furthermore the computation of the optimal control input generally requires a backward integration or recursion. Instead of struggling with this difficulty, one can restrict the stimulus $I$ to a well known natural form such as sum of phased cosines as shown below:
\begin{equation}
	 I=\sum_{n=1}^{N}A_n \cos\left(\omega_{n}t+\phi_{n}\right)\label{eq:cosine-stimulus}
\end{equation}
where $A_n$ is the amplitude, $\omega_n$ is the frequency of the $n$-th Fourier component, and $\phi_n$ is the phase of the component. 
We choose a base frequency ${\omega_1}$ and set the frequencies of all other components as the harmonics: $\omega_n=n\omega_1$ for $n=1,\ldots, N$. 
Now the stimulus parameters can be summarized by the stimulus parameter vector:
\begin{equation}
	{\bf x}=[A_1,\ldots,A_{N}, \phi_1,\ldots, \phi_{N}]
	\label{eq:x}
\end{equation}
We sometimes refer to $\bf x$ as the stimulus, with it understood that it really means a set of parameters that uniquely specify the actual stimulus $I$.

Some popular choices of the objective function are based on the Fisher information matrix, which is generally defined as:
\begin{equation}
 	F_{ij}\left({\bf x},\theta\right)=\left<\frac{\partial \ln p({\bf r} \mid {\bf x},\theta)}{\partial\theta_i}
	\frac{\partial \ln p({\bf r} \mid {\bf x},\theta)}{\partial\theta_j}\right>
	\label{eq:Fisher-information-matrix}
\end{equation}
where $p({\bf r} \mid {\bf x},\theta)$ is the probability distribution of the response $\bf r$ to a given stimulus $\bf x$. \textcolor{red}{One should here note that, the term $ p({\bf r} \mid {\bf x},\theta) $ is different from the terms $ p\left(t_{1},t_{2},\ldots,t_{K}\right) $ in \eqref{eq:lkl-enbrown} and $ p\left(S_m\mid\theta\right) $ in \eqref{eq:pSm}. The latter two are derived from the spiking likelihood function presented in \cite{uteden2008pointprocesses,brown2002time} whereas the definition $ p({\bf r} \mid {\bf x},\theta) $ is derived from \eqref{eq:inhomogeneous-poisson}. However only the related results are provided in this text in order to save space. 
In \eqref{eq:Fisher-information-matrix}, $\left<\;\right>$ represents the weighted average over all possible responses $\bf r$ to all available sets of stimuli $\bf x$. This average is calculated based on the response probability distribution $ p({\bf r} \mid {\bf x},\theta) $. Stimulus variable $\bf x$ can be represented by a group of parameters such as the Fourier series parameters in \eqref{eq:cosine-stimulus} which are $ A_n,\ \omega_n $ and $ \phi_n $. The frequency parameter may or may not be fixed depending on the requirements or computational burden.}   

The Fisher information matrix reflects the amount of information contained in the noisy response ${\bf r}$ about the model parameters  ${\theta}$, assuming a generative model given by the conditional probability $p({\bf r} \mid {\bf x},\theta)$.  So the stimulus designed by maximizing a certain measure of the Fisher information matrix \eqref{eq:Fisher-information-matrix} is expected to decrease the error of the estimation of the parameters ${\theta}$.

The utility function $U$ can be chosen as a scalar function of the Fisher information matrix $\bf F$. 
One theoretically well founded popular choice is the D-optimal design:
\begin{equation}
	U({\bf x},\theta)=\det {\bf F}({\bf x},\theta)
	\label{eq:Dopt}
\end{equation}
although the determinant of the Fisher information matrix is not always easy to optimize.
The A-optimal design is based on the trace of the Fisher information matrix and is easier to optimize:
\begin{equation}
	U({\bf x},\theta)= {\rm tr\,} {\bf F}({\bf x},\theta)
	\label{eq:Aopt}
\end{equation}
Another alternative is the E-optimal design where the objective function is the smallest eigenvalue of the Fisher information matrix.
In this paper the A-optimality measure of the information matrix is preferred. There is an obvious reason for this preference. As the computational complexity of the optimization algorithms are expected to be high, the necessity of numerical derivative computation should be avoided as much as possible. Since it is not easy to evaluate the derivatives of the eigenvalues and determinants by any means other than numerical approximations it will be convenient to apply a criterion like A-optimality which has a direct relationship like the sums of the diagonal elements.

In the beginning of this section we mentioned that an optimally design stimulus is expected to depend on which parameter is supposed to be \textcolor{red}{estimated}. 
Since a scalar utility function in \eqref{eq:Dopt} or \eqref{eq:Aopt} depends on all the parameters 
$\theta=[\theta_1,\theta_2,\ldots]$,
optimizing a single scalar function is sufficient to recover all the parameters. 
When a sequence of  stimuli are generated by optimal design, the stimuli may sometimes alternate spontaneously as if the optimization is performed with respect to each of the parameters one by one \citep{dimattina2011active}.

In our network model, the recorded spike train has an inhomogeneous Poisson distribution with the rate function ${r_e\left(t\right)}$. We write this rate as $r_e\left(t,{\bf x},\theta\right)$ to emphasize that it is a time-varying function that depends on both the stimulus $\bf x$ and the network parameters $\bf x$.
For a small time window of duration $\Delta t$ and centered at time $t$, the Fisher information matrix entry in \eqref{eq:Fisher-information-matrix} is reduced to:
\begin{equation}
	 F_{ij}\left(t, {\bf x},\theta\right)
	 =\frac{\Delta t}{r_e\left(t,{\bf x},\theta\right)}\frac{\partial{r_e}\left(t,{\bf x},\theta\right)}{\partial\theta_i}\frac{\partial{r_e}\left(t,{\bf x},\theta\right)}{\partial\theta_j}
	 \label{eq:fisher-information-over-poisson}
\end{equation}  
Since the Poisson rate function $r_e$ varies with time, the A-optimal utility function in \eqref{eq:Aopt} should be modified by including integration over time:
 \begin{equation}
 	U\left({\bf x},\theta\right)=\int_0^{T} {\rm tr\,} {\bf F}(t,{\bf x},\theta) dt
	=\int_0^T \sum_{k=1}^8\frac{1}{r_e(t,{\bf x},\theta)}\left(\frac{\partial r_e(t,{\bf x},\theta)}{\partial \theta_k}\right)^2 dt \\
	\label{eq:oed-objective-general}
\end{equation}
Here the time window $\Delta t$ is ignored because it is a constant coefficient that does not affect the result of the optimization.

For convenience, we can also define the objective function with respect to a single parameter $\theta_k$ as follows:
 \begin{equation}
 	U_k\left({\bf x},\theta\right)=\int_0^T \frac{1}{r_e(t,{\bf x},\theta)}\left(\frac{\partial r_e(t,{\bf x},\theta)}{\partial \theta_k}\right)^2 dt \\ \label{eq:fisher-wrto-single-param}
\end{equation}
The objective function in \eqref{eq:oed-objective-general} is identical to $\sum_{k=1}^8 U_k$.

The optimization of the D-optimal criterion in \eqref{eq:Dopt} is not affected by parameter rescaling, or changing the units of parameters. 
For example, changing the unit of parameter 1 (say, from msec$^{-1}$ to sec$^{-1}$) is equivalent to rescaling the parameter by a constant coefficient: $\theta_1 \to c \theta_1$. 
The effect of this transformation is equivalent to a rescaling of the determinant of the Fisher information matrix by a constant, namely, $\det{\bf F}\to (\det{\bf F})/c^{16}$, which does not affect the location of the maximum of \eqref{eq:Dopt}.  
By contrast, the criterion function in \eqref{eq:Aopt} or \eqref{eq:oed-objective-general} are affected by parameter rescaling. 
A parameter with a smaller unit would tend to have larger derivative value  
and therefore contribute more to \eqref{eq:oed-objective-general} than a parameter with a large unit.
To alleviate this problem, we use $U_k$ one by one to generate the stimuli. That is, stimulus 1 is generated by maximizing $U_1$, and stimulus 2 is generated by maximizing $U_2$, and so on. Once the 8th stimulus is generated by maximizing $U_8$, we go back and use $U_1$ to generate the next stimulus, and so on. 
Finally, an alternative way to get rid of scale dependence is to introduce logarithm and use $U=\sum_{k}\ln U_k$ as the criterion, which, however, may become degenerate when $U_k$ approaches 0.
\subsection{Gradient Computation} \label{sub:grad-computation}

As just explained in the previous section about the computational issues in this research, the gradient computation decreases the computation durations considerably. The main issue with this fact is the lack of closed form expressions like in the case of static non-linear mappings as the model. In researches such as \citep{flila2010combined}, \citep{telen2012approximate} and \citep{telen2012robust} the gradients are computed as a self contained differential equation which is formed by taking the derivatives of the model equations \eqref{eq:2-exc-inh-network-a} and \eqref{eq:2-exc-inh-network-b} from both sides. 

Compiling all the information in this section one can write the gradient of the Fisher Information Measure (i.e. the Fisher Information Matrix with a certain optimality criterion such as A-Optimality). In the beginning of this section, it is stated that the sensitivity levels of the firing rate w.r.to different network parameters are different and thus it would be convenient to maximize the fisher information for a single parameter at a time.

\noindent The optimization as expressed in \eqref{eq:oed-general-def} the optimal design problem is converted into a parameter optimization problem to optimize the amplitudes ${A_n}$'s and ${\phi_n}$'s of the stimulus. For the sake of simplicity and modularity in programming, these equations can be written using their shorthand notations.

\noindent Let
\begin{equation}
	{\bf x}=[x_1,x_2,\ldots, x_{2N}]=[A_1,\ldots,A_{N}, \phi_1,\ldots, \phi_{N}]
	\label{eq:param-stimulus}
\end{equation}

\noindent We write the state of the network as a vector:
$ {\bf v}=[V_e, V_i]^{\rm T} $

\noindent and we need derivatives such as $\frac{d}{dt}\frac{\partial {\bf v}}{\partial \bf x}$. 
Here, the idea is to use these derivatives as variables that can be solved directly from differential equations derived from the original dynamical equations in \eqref{eq:our-model-matrix-form}.

\noindent For the optimal design, one can write the following:
\begin{equation}
	 \frac{d}{dt}\frac{\partial {\bf v}}{\partial \bf x}
	 =\left[\begin{array}{cc}
		\beta_{e} & 0\\
		0 & \beta_{i}
	\end{array}\right]
	\left\{ -\frac{\partial \bf v}{\partial \bf x}+\left[\begin{array}{cc}
	w_{ee} & -w_{ei}\\
	w_{ie} & -w_{ii}
	\end{array}\right]
	\left[\begin{array}{cc}
	 g'_{e}\left(V_{e}\right) & 0\\
	0 &  g'_{i}\left(V_{i}\right)
	\end{array}\right]\frac{\partial \bf v}{\partial \bf x}+\left[\begin{array}{c}
	w_{e}\\
	w_{i}
	\end{array}\right]\frac{\partial I}{\partial \bf x}\right\} 
	\label{eq:intermediate-step-for-DVU}
\end{equation}
where $g'_e$ and $g'_i$ are the derivatives of the gain functions $g_e$ and $g_i$ respectively,  and the matrices derivatives are defined in the usual manner:
\begin{equation}
	 \frac{\partial I}{\partial \bf x}=\left[\begin{array}{cccccc}
	\displaystyle\frac{\partial I}{\partial A_{1}}  & \ldots & \displaystyle\frac{\partial I}{\partial A_{N}},
	 &\displaystyle \frac{\partial I}{\partial\phi_{1}}  & \ldots  & \displaystyle\frac{\partial I}{\partial\phi_{N}}\end{array}\right]
\end{equation}
and
\begin{equation}
	\frac{\partial \bf v}{\partial \bf x}= \left[\begin{array}{cccccc}\displaystyle
	\frac{\partial V_{e}}{\partial A_{1}} & \ldots & \displaystyle\frac{\partial V_{e}}{\partial A_{N}} & \displaystyle\frac{\partial V_{e}}{\partial\phi_{1}}  
	& \ldots & \displaystyle\frac{\partial V_{e}}{\partial\phi_{N}}\\ 
	\mbox{}\\
	\displaystyle\frac{\partial V_{i}}{\partial A_{1}}  & \ldots &\displaystyle \frac{\partial V_{i}}{\partial A_{N}} 
	& \displaystyle \frac{\partial V_{i}}{\partial\phi_{1}} & \ldots & \displaystyle\frac{\partial V_{i}}{\partial\phi_{N}}
\end{array}\right]\label{eq:potential-gradient-stimulus-parameters}
\end{equation}

\noindent \eqref{eq:intermediate-step-for-DVU} can be written equivalently in the shorthand form:
\begin{equation}
	\frac{d}{dt}\frac{\partial \bf v}{\partial \bf x}
	={\bf B}\left\{ -\frac{\partial \bf v}{\partial \bf x}+{\bf WG}\frac{\partial \bf v}{\partial \bf x}+{\bf w} \frac{\partial I}{\partial \bf x}\right\}
	 \label{eq:gradient-evolution-wrt-stimulus-param}
\end{equation}
where
${{\bf B}=\left[\begin{array}{cc}
	\beta_e & 0 \\ 
	0 & \beta_i
	\end{array}\right]}$, 
${{\bf W}=\left[\begin{array}{cc}
	w_{ee} & -w_{ei} \\ 
	w_{ie} & -w_{ii}
	\end{array}\right] }$,
${\bf G}=\left[\begin{array}{cc}
	 g'_{e}(V_{e}) & 0\\
	0 & g'_{i}(V_{i})
	\end{array}\right]$, 
and
${{\bf w}=\left[\begin{array}{c}
	w_e \\ 
	w_i
	\end{array}\right]}$.

\noindent In the evaluation of the Fisher Information Matrix and the gradients in estimation one will need the derivatives with respect to the parameters 
${\theta}=[\theta_1,\ldots,\theta_8]=\left[\beta_e,\beta_i,w_e,w_i,w_{ee},w_{ei},w_{ie},w_{ii}\right]$.

\noindent It follows from the original dynamical equation in \eqref{eq:our-model-matrix-form} that    
\begin{equation}
	\frac{d}{dt}\frac{\partial \bf v}{\partial \theta_k}={\bf B}\left\{ -\frac{\partial \bf v}{\partial \theta_k}
	+{\bf WG}\frac{\partial \bf v}{\partial \theta_k}\right\} +{\bf z}_k
	\label{eq:core-components-compact}
\end{equation}
where
$\displaystyle{\frac{\partial \bf v}{\partial \theta_k}
	=\left[
	\frac{\partial V_{e}}{\partial\theta_k},
	\frac{\partial V_{i}}{\partial\theta_k}
	\right]^{\rm T}}$
and the last term ${\bf z}_k$ refers to the extra components resulting from the chain rule of differentiation. 
These extra terms are presented in \textbf{Table \ref{Tab:ext-comp-derivat}}.

\begin{table}[H]
\centering
\caption{The extra components ${\bf z}_k$ in Equation \eqref{eq:core-components-compact} }\label{Tab:ext-comp-derivat}
\begin{tabular}{ccc}
\toprule
$k$ & Parameter $\theta_k$ & Extra term ${\bf z}_k$  in Equation (\ref{eq:core-components-compact}) \\
\midrule
1 & $\beta_{e}$ & $\left[\begin{array}{cc}
	1 & 0\\
	0 & 0
	\end{array}\right]\left\{ -\left[\begin{array}{c}
	V_{e}\\
	V_{i}
	\end{array}\right]+
	{\bf W}\left[\begin{array}{c}
	g_{e}\left(V_{e}\right)\\
	g_{i}\left(V_{i}\right)
	\end{array}\right]+{\bf w}I\right\} $ \\
	\midrule
2& $\beta_{i}$ & $\left[\begin{array}{cc}
	0 & 0\\
	0 & 1
	\end{array}\right]\left\{ -\left[\begin{array}{c}
	V_{e}\\
	V_{i}
	\end{array}\right]+{\bf W}\left[\begin{array}{c}
	g_{e}\left(V_{e}\right)\\
	g_{i}\left(V_{i}\right)
	\end{array}\right]+{\bf w}I\right\} $ \\  
	\midrule
3& $w_{e}$ & ${\bf B}\left[\begin{array}{c}
	1\\
	0\\
	\end{array}\right]I$\\  
	\midrule
4& $w_{i}$ & ${\bf B}\left[\begin{array}{c}
	0\\
	1\\
	\end{array}\right]I$\\   
	\midrule 
5& $w_{ee}$ & ${\bf B} \left[\begin{array}{cc}
	1 & 0\\
	0 & 0
	\end{array}\right]\left[\begin{array}{c}
	g_{e}\left(V_{e}\right)\\
	g_{i}\left(V_{i}\right)
	\end{array}\right] $ \\ 
	\midrule
6& $w_{ei}$ & ${\bf B} \left[\begin{array}{cc}
	0 & -1\\
	0 & 0
	\end{array}\right]\left[\begin{array}{c}
	g_{e}\left(V_{e}\right)\\
	g_{i}\left(V_{i}\right)
	\end{array}\right]$ \\ 
	\midrule
7& $w_{ie}$ & ${\bf B} \left[\begin{array}{cc}
	0 & 0\\
	1 & 0
	\end{array}\right]\left[\begin{array}{c}
	g_{e}\left(V_{e}\right)\\
	g_{i}\left(V_{i}\right)
	\end{array}\right] $ \\ 
\midrule
8& $w_{ii}$ & ${\bf B}
	 \left[\begin{array}{cc}
	0 & 0\\
	0 & -1
	\end{array}\right]
	\left[\begin{array}{c}
	g_{e}\left(V_{e}\right)\\
	g_{i}\left(V_{i}\right)
	\end{array}\right]$  \\ 
\bottomrule
\end{tabular} 
\end{table}

The maximization of the Fisher Information Measure in trace form requires its gradients and they involve second order cross derivatives of the membrane potentials with respect to parameters ${\left(\theta_k\right)}$ and stimulus parameters ${\left(I_l\right)}$. 

\noindent Taking derivative of \eqref{eq:gradient-evolution-wrt-stimulus-param} with respect to $\theta_k$, we find:
\begin{equation}
	 \frac{d}{dt}\frac{\partial^2 \bf v}{\partial {\bf x}\partial\theta_k}
	 ={\bf B}\left\{ -\frac{\partial^2 \bf v}{\partial {\bf x}\partial\theta_k}
	 +{\bf WG}\frac{\partial^2 \bf v}{{\partial\bf x}\partial\theta_k}+{\bf WG'}\,{\rm diag}\! \left(\frac{\partial \bf v}{\partial\theta_k}\right)\frac{\partial \bf v}{\partial \bf x}\right\} 
	 +{\bf Z}_k \label{eq:membrane-potential-derivative-second-order}
\end{equation}
where 
${\bf G}'=\left[\begin{array}{cc}
	g''_{e}(V_e) & 0\\
	0 & g''_{i}(V_i)
	\end{array}\right]$, 
$\displaystyle{\rm diag}\!\left(\frac{\partial \bf v}{\partial\theta_k}\right)=\left[\begin{array}{cc}
	\frac{\partial V_{e}}{\partial\theta_k} & 0\\
	0 & \frac{\partial V_{i}}{\partial\theta_k}
	\end{array}\right]$,
and 
\begin{equation}
 	\frac{\partial^2 \bf v}{\partial{\bf  x}\partial\theta_k}=\left[\begin{array}{cccccc}
	\displaystyle\frac{\partial^{2}V_{e}}{\partial A_{1}\partial\theta_k} &  \ldots 
	& \displaystyle \frac{\partial^{2}V_{e}}{\partial A_{N}\partial\theta_k} & \displaystyle\frac{\partial^{2}V_{e}}{\partial\phi_{1}\partial\theta_k} 
	 & \ldots & \displaystyle\frac{\partial^{2}V_{e}}{\partial\phi_{N}\partial\theta_k}\\
	 \mbox{}\\
	\displaystyle\frac{\partial^{2}V_{i}}{\partial A_{1}\partial\theta_k} & \ldots & \displaystyle\frac{\partial^{2}V_{i}}{\partial A_{N}\partial\theta_k} 
	& \displaystyle\frac{\partial^{2}V_{i}}{\partial\phi_{1}\partial\theta_k} &  \ldots & \displaystyle \frac{\partial^{2}V_{i}}{\partial\phi_{N}\partial\theta_k}
\end{array}\right]\label{eq:DVWU}
\end{equation}
which is compatible with \eqref{eq:potential-gradient-stimulus-parameters}.
The last term ${\bf Z}_k$ is specified in Table \ref{Tab:ext-comp-2nd-derivat}

\begin{table}[H]
\centering
\caption{The extra components ${\bf Z}_k$ in Equation \eqref{eq:membrane-potential-derivative-second-order}} 
\label{Tab:ext-comp-2nd-derivat}
\begin{tabular}{ccc}
\toprule
$k$ & Parameter $\theta_k$ & Extra term ${\bf Z}_k$  in Equation (\ref{eq:membrane-potential-derivative-second-order}) \\
\midrule
1& $\beta_{e}$ & $\left[\begin{array}{cc}
	1 & 0\\
	0 & 0
	\end{array}\right]
	\displaystyle\left\{ -\frac{\partial \bf v}{\partial \bf x}+{\bf WG}\frac{\partial \bf v}{\partial \bf x}+{\bf w} \frac{\partial I}{\partial \bf x} \right\} $ \\ 
\midrule
1& $\beta_{i}$ & $\left[\begin{array}{cc}
	0 & 0\\
	0 & 1
	\end{array}\right]
	\displaystyle\left\{ -\frac{\partial \bf v}{\partial \bf x}+{\bf WG}\frac{\partial \bf v}{\partial \bf  x}+{\bf w} \frac{\partial I}{\partial \bf x} \right\} $ \\ 
\midrule 
3 & $w_{e}$ & ${\bf B}\left[\begin{array}{c}
	1\\
	0
	\end{array}\right]
	\displaystyle\frac{\partial I}{\partial \bf x}$ \\ 
\midrule
4&  $w_{i}$ & ${\bf B}\left[\begin{array}{c}
	0\\
	1
	\end{array}\right]
	\displaystyle\frac{\partial I}{\partial \bf x}$ \\ 
\midrule
5& $w_{ee}$ & ${\bf B}\left[\begin{array}{cc}
	1 & 0\\
	0 & 0
	\end{array}\right]
	\displaystyle 
	{\bf G}\frac{\partial \bf v}{\partial \bf x}$ \\ 
\midrule
6& $w_{ei}$ & ${\bf B}\left[\begin{array}{cc}
	0 & -1\\
	0 & 0
	\end{array}\right]
	\displaystyle {\bf G}\frac{\partial \bf  v}{\partial \bf x}$ \\ 
\midrule
7& $w_{ie}$ & ${\bf B}\left[\begin{array}{cc}
	0 & 0\\
	1 & 0
	\end{array}\right]
	\displaystyle {\bf G}\frac{\partial \bf v}{\partial \bf x}$ \\ 
\midrule
8& $w_{ii}$ & ${\bf B}\left[\begin{array}{cc}
	0 & 0\\
	0 & -1
	\end{array}\right]
	\displaystyle {\bf G}\frac{\partial \bf  v}{\partial \bf x}$ \\ 
\bottomrule
\end{tabular}
\end{table}      

To compute the gradients needed for optimizing the objective function based on Fisher Information or for performing  maximum-likelihood estimation, 
we need to evaluate the derivatives of 
the mean firing rate  ${r_e=g_e\left(V_e\right)}$ with respect to the network parameters $\theta_k$ in \eqref{eq:theta-ctrnn-param} or the stimulus parameters $x_j$ in \eqref{eq:param-stimulus}. 
The first and the second order derivatives are:
\begin{equation}
	 \frac{\partial{r_e}}{\partial{\theta_k}}=g'_e \left(V_e\right) \frac{\partial{V_e}}{\partial{\theta_k}}
	 \label{eq:derivat-firing-rate-potential-theta}
\end{equation}
\begin{equation}
	 \frac{\partial{r_e}}{\partial{x_l}}=g'_e\left(V_e\right)\frac{\partial{V_e}}{\partial{x_l}}
	 \label{eq:derivat-firing-rate-potential-U}
\end{equation}
\begin{equation}
	\frac{\partial^{2}r_{e}}{\partial x_{l}\partial\theta_k}
	=g''_{e}\left(V_{e}\right)\frac{\partial V_{e}}{\partial x_{l}}\frac{\partial V_{e}}{\partial\theta_k}
	+g'_{e}\left(V_{e}\right)\frac{\partial^{2}V_{e}}{\partial x_{l}\partial\theta_k}
	\label{eq:2nd-derivat-rate-potential}
\end{equation}
These formulas are expressed in terms of the derivatives $\frac{\partial{V_e}}{\partial{\theta_k}}$, $\frac{\partial{V_e}}{\partial{x_l}}$ and $\frac{\partial^{2}V_{e}}{\partial x_{l}\partial\theta_k}$, which are regarded as dynamical variables that can be solved 
from the three differential equations \eqref{eq:gradient-evolution-wrt-stimulus-param}--\eqref{eq:membrane-potential-derivative-second-order}.
Here the initial conditions were always assumed to be the equilibrium state. The initial values of the derivatives can be set to zero as recommended in \citep{telen2012approximate}. 

\noindent The gradient of the Fisher Information Measure in \eqref{eq:oed-objective-general} with respect to the stimulus parameters can be written as
\begin{eqnarray}
	\frac{\partial U}{\partial x_{l}}&=&\int_0^T \frac{\partial }{\partial x_{l}}\sum_{k=1}^8  {\frac{1}{r_e}\left(\frac{\partial{r_e}}{\partial{\theta_k}}\right)^2}dt\\
	&=& \int_0^T \sum_{k=1}^8 \left\{ -\frac{1}{r_{e}^{2}}\frac{\partial r_{e}}{\partial x_{l}}\left(\frac{\partial r_{e}}{\partial\theta_k}\right)^{2}
	+\frac{2}{r_{e}}\frac{\partial r_{e}}{\partial\theta_k} \frac{\partial^{2}r_{e}}{\partial x_{l}\partial\theta_k}\right\}dt
	\label{eq:grad-fisher-normal}
\end{eqnarray} 
which the last expression is written in terms of the derivatives that are already evaluated by equations \eqref{eq:derivat-firing-rate-potential-theta}--\eqref{eq:2nd-derivat-rate-potential}. If the Fisher Information is computed with respect to one parameter at a time as shown in \eqref{eq:fisher-wrto-single-param}, one rewrite the above by removing the summation as:

\begin{eqnarray}
	\frac{\partial U_k}{\partial x_{l}}&=&\int_0^T \frac{\partial }{\partial x_{l}} {\frac{1}{r_e}\left(\frac{\partial{r_e}}{\partial{\theta_k}}\right)^2}dt\\
	&=& \int_0^T \left\{ -\frac{1}{r_{e}^{2}}\frac{\partial r_{e}}{\partial x_{l}}\left(\frac{\partial r_{e}}{\partial\theta_k}\right)^{2}
	+\frac{2}{r_{e}}\frac{\partial r_{e}}{\partial\theta_k} \frac{\partial^{2}r_{e}}{\partial x_{l}\partial\theta_k}\right\}dt
	\label{eq:grad-fisher-single}
\end{eqnarray} 

Lastly, for maximum likelihood estimation, one need the gradient of the log likelihood function of spike trains in \eqref{eq:complete-likelihood-compact}:
\begin{equation}
	\frac{\partial{l}}{\partial{\theta_k}}
	=-\sum_{m=1}^{M}\int_{0}^{T}\frac{\partial{r_e^{(m)}}\!\left(t\right)}{\partial{\theta_k}}\!\left(t\right)dt 
	+\sum_{m=1}^{M}\sum_{k=1}^{K_m}{{r_e^{(m)}}\!\left(t_{k}^{(m)}\right)}^{-1}\frac{\partial{r_e^{(m)}}\!\left(t_{k}^{(m)}\right)}{\partial{\theta_k}} 
 	\label{eq:derivative-log-likelihood-compact}
\end{equation}
which is written in terms of derivatives that can be evaluated by \eqref{eq:derivat-firing-rate-potential-theta}. 

\subsection{Other Numerical Issues Related to Optimization} \label{sub:other-numerical}

\indent Up to this point, the theoretical grounds of this research are presented. In order to achieve the results one needs two separate maximization algorithms targeting \eqref{eq:mle-arg-max-log} and \eqref{eq:oed-general-def}. Optimization algorithms have varieties in certain aspects. One of these classifications is the gradient requirements. The global optimization algorithms such as genetic algorithms (MATLAB\textsuperscript{\circledR} \emph{ga}), simulated annealing (MATLAB\textsuperscript{\circledR} \emph{simulannealbnd}) and pattern search (MATLAB\textsuperscript{\circledR} \emph{patternsearch}) do not require gradient computations. However due to their stochastic nature, they do a lot of computations which increase the computation duration drastically. In addition, that stochastic nature leads to different results at the end of different runs. Disabling this randomness, might have adverse effects on the optimization performance and their ability of finding a global optimum solution. \textcolor{red}{In order not to be trapped in these issues we will stick to gradient based optimization routines such as constrained interior-point gradient descent. MATLAB\textsuperscript{\circledR} provides these algorithms through its optimization toolbox. Interior point and similar methods are available in \emph{fmincon} function.} We can write the following facts about the optimization algorithm:

\begin{itemize}
\item MATLAB\textsuperscript{\circledR}'s \emph{fmincon} allows the user to set constraints on the solution.In the optimal design the stimulus amplitudes ${A_n}$ will have an upper bound (numerical details will be discussed \textbf{Section \ref{sub:statistics-numerical-discuss}}. The stimulus amplitude is expected to tend to the upper bound. The parameter estimation process will also be benefited from similar bounds as they are assigned to be positive in \eqref{eq:our-model-matrix-form}. In addition assignment of bounds on parameters will aid in prevention of overstimulation of the model.  
\item \emph{fmincon} and similar algorithms are local optimizers. In order to find a good optimum, the algorithms are often repeated with multiple initial guesses and the best one is chosen according to the value of the objective and gradient value at the termination point. This is especially important in the optimal design part as Fisher Information measure in \eqref{eq:oed-objective-general} may have lots of local minima. The same factors do exist for likelihood function \eqref{eq:complete-likelihood-compact} however as the number of samples ${M}$ increases the likelihood function tend to converge to the same optimum for different initial guesses. See \textbf{Section \ref{sub:local-maxima-discussion}} for a detailed discussion of this issue. 
\item The local optimization algorithms need the gradient of the objective functions. In MATLAB\textsuperscript{\circledR}'s \emph{fmincon} and other similar packages have the ability to compute the gradients numerically. However this capability may not be an advantage in this sort of complicated problems for two main reasons. First of all, numerical gradient computation increases the risk of singularities in the solution. Secondly, the computational duration increases dramatically. It is often noted that, the numerical differentiation in \emph{fmincon} at least doubles the computational duration. This is unacceptable in the context of this research.        
\item Longer simulation times due to the computational complexity of the overall problem can be eliminated to a certain extent (about the 5 times shorter duration) by utilizing the MATLAB\textsuperscript{\circledR}'s parallel computation facilities (such as replacing standard \emph{for} loops by \emph{parfor} loops).   
\item The MATLAB\textsuperscript{\circledR} version used in this research is R2013b and the \emph{fmincon} version included in this package has the interior-point method of constrained non-linear optimization by default. This is not the only method available in \emph{fmincon} and it can always be modified by changing the algorithm option. However, the interior-point method appears to be the most efficient among all available options.            
\end{itemize}

\subsection{Procedural Information} \label{sub:procedure-info}

In this section, we will summarize the overall procedure to give an insight about how the optimal design and parameter estimation algorithm works together in a automatized loop. Before giving a start, it is worth to discuss the initial parameter problem. In the beginning of an experiment (or simulation in our case), one usually has no idea about the true or approximate values of the network parameter vector $\theta$. However, the optimal design always needs a current estimate. Thus one should assign a randomly chosen initial parameter vector. A good rule to get that value is to draw one sample from a parametric subspace of which members are uniformly distributed between the lower and upper bounds of the parameters. One can find the details about the parametric bounds in \textbf{Table \ref{Tab:true-values-and-statistics}} \textcolor{red}{Setting a parametric bound on the stimulus amplitude coefficients ($ A_n $ in \eqref{eq:cosine-stimulus}) might be critical. Too large amplitudes may lead to instabilities or very large responses which may break the optimal design procedures. Because of this fact, an upper bound on $ A_n $'s will most often required. A few number of initial simulations will be required to determine a nominal value for those bounds. If no instabilities are detected one will be fine with the decided value of the bounds on $ A_n $ (which is termed as $ A_{\max} $) and can continue the optimization with these bounds. Mostly, the optimization procedure will fail when the assigned value of the bounds are too large. }

The algorithmic details of the overall process are summarized as shown below:

\begin{enumerate}
\item Set $i=1$
\item Generate a random parameter vector $\theta^0$ from a uniform distribution between $\theta_{min}$ and $\theta_{max}$ (from \textbf{Table \ref{Tab:true-values-and-statistics}}). 
\item Set $\hat\theta=\theta^0$
\item Set $k=1$ \label{enum:oed-step-1} 
\item Optimize $U_k$ based on $\hat\theta$ to generate stimulus $\bf x$ \label{enum:oed-step-k}
\item Using the stimulus $\bf x$ perform a maximum likelihood estimation to find a new estimate of $\theta$ and replace $\hat\theta$ by the new estimate. 
\item Set $k\rightarrow{k+1}$ and go to \textbf{Step \ref{enum:oed-step-k}} . If $k>8$ set $i\rightarrow{i+1}$  
\item If $i>N_{itr}$ stop and report the result as $\hat\theta^{final}=\hat\theta$ otherwise go to \textbf{Step \ref{enum:oed-step-1}}
  
\end{enumerate} 

\noindent The results related to the application of all theoretical and algorithmic developments in the \textbf{Sections \ref{sub:ctrnn-theory}} to \textbf{\ref{sub:procedure-info}} will be presented and discussed in \textbf{Sections \ref{sub:results}} and \textbf{\ref{sub:discussion}}.

\section{Results} \label{sub:results}
In this section, we will summarize the functional and numerical details of the combined optimal design - parameter estimation algorithm and present the results in comparison with the random stimuli tests. This section is divided into several sections discussing the numerical details of an example application, statistical properties of the optimal stimuli, accuracy of the parameter estimation and parameter confounding phenomenon.  

\subsection{Details of the example problem} \label{sub:example-details}
This section is devoted to the detailed presentation of the simulation set-up. An numerical example will be presented which will demonstrate our optimal design approach. In the example application, the algorithms  presented in \textbf{Sections \ref{Sub:jmle-theory}} and \textbf{\ref{Sub:oed-theory}} are applied to probe an EI network.
 
\noindent In order to verify the performance of the parameter estimation we have to compare the estimates with their true values. So we will need a set of reference values of the model parameters in \eqref{eq:our-model-matrix-form}. These are shown in \textbf{Table \ref{Tab:true-values-and-statistics}}. 

\begin{table}[H]
\centering
\scriptsize
\caption{ 
The true values, lower and upper bounds of parameters of model in \eqref{eq:our-model-matrix-form}. Mean values and standard deviations of the estimates for the case $M=120$ are also shown for convenience.  
} \label{Tab:true-values-and-statistics}
\begin{tabular}{ccccccccc}
    \toprule 
    Parameter & ${\beta_e}$ ($ \nicefrac{1}{s} $) & ${\beta_i}$ ($ \nicefrac{1}{s} $) & ${w_e}$ (k$ \Omega $) & ${w_i}$ (k$ \Omega $) & ${w_{ee}}$ (mV$ \cdot $s) & ${w_{ei}}$ (mV$ \cdot $s) & ${w_{ie}}$ (mV$ \cdot $s) & ${w_{ii}}$ (mV$ \cdot $s) \\ 
    \midrule
    True value $\left(\theta\right)$ & ${50}$ & ${25}$ & ${1.0}$ & ${0.7}$ & ${1.2}$ & ${2.0}$ & ${0.7}$ & ${0.4}$ \\ 
    \midrule
    Lower bound $\left(\theta_{min}\right)$ & 0 & 0 & 0 & 0 & 0 & 0 & 0 & 0 \\
    \midrule
    Upper bound $\left(\theta_{max}\right)$ & 100 & 100 &2 &2 & 3 & 3 & 3 & 3\\ 
    \midrule
    Mean (OED) & $49.9362$ & $25.1842$ & $1.0016$ & $0.6975$ & $1.2267$ & $2.0690$ & $0.7120$ & $0.4562$\\
    \midrule
    STD (OED)  & $0.8620$  & $1.4373$  & $0.0357$ & $0.0678$ & $0.0690$ & $0.1872$ & $0.0949$ & $0.1987$\\
    \midrule
    Mean (RAND) & $50.0685$ & $25.2265$ & $0.9979$ & $0.7149$ & $1.2469$ & $2.0714$ & $0.7844$ & $0.5694$\\
    \midrule
    STD (RAND) & $1.6271$ & $1.9601$ & $0.0394$ & $0.0996$ & $0.1066$ & $0.2271$ & $0.1708$ & $0.3735$\\
    \bottomrule 
\end{tabular}    
\end{table}

Our model in \eqref{eq:our-model-matrix-form} has two more important components which are the gain functions $g_e\left(V_e\right)$ and $g_i\left(V_i\right)$. These are obtained by setting $j$ in \eqref{eq:sigmoid-general} by either '$e$' or '$i$'. So one has $6$ additional parameters $\left[\Gamma_e,a_e,h_e,\Gamma_i,a_i,h_i\right]$ which have direct effect on the neural model behaviour. This research targeted the estimation of the network parameters only. Because of that, the parameters of the gain functions are kept as fixed and they have the values $\Gamma_e=100,a_e=0.04,h_e=70,\Gamma_i=50,a_i=0.04,h_i=35$. 

\begin{figure}[H]
\centering
\includegraphics[scale=0.7]{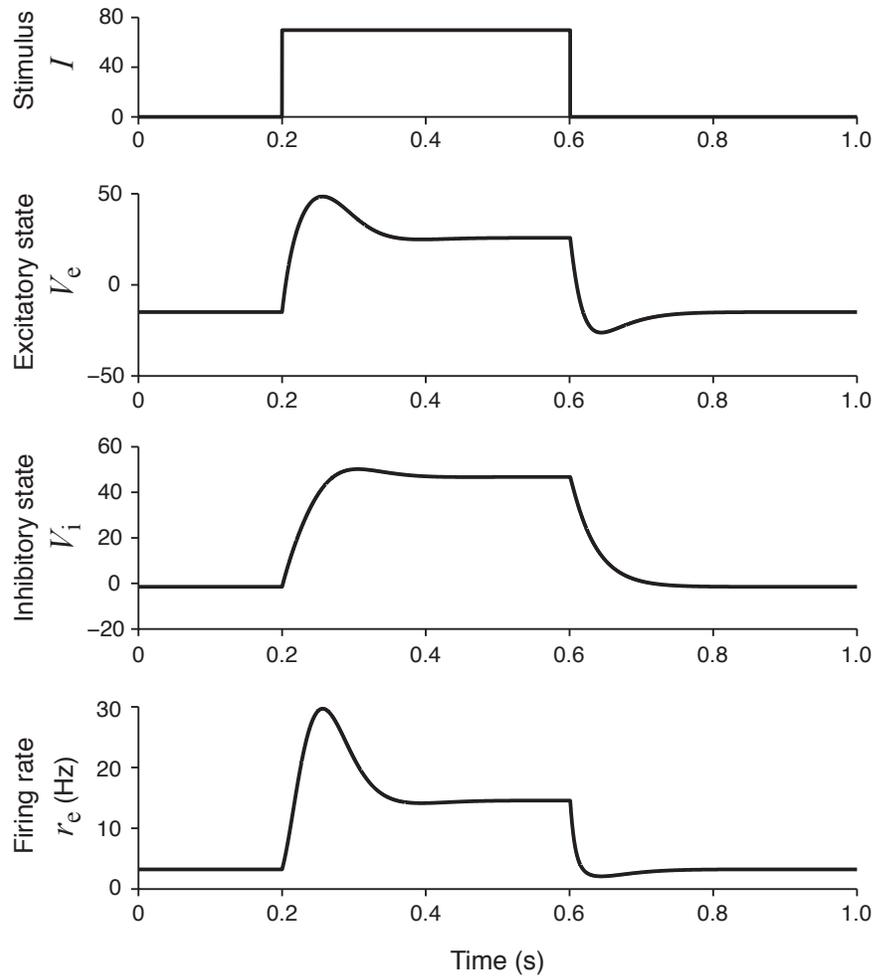}
\caption{ The network model in \textbf{Figure \ref{fig:generic-ctrnn}B} in response to a square-wave stimulus. The states of the excitatory and inhibitory units, $V_e$ and $V_i$, are shown, together with the continuous firing rate of the excitatory unit, $r_e=g_e\left(V_e\right)$. The firing rate of the excitatory unit (bottom panel) has a transient component with higher firing rates, followed by a plateau or sustained component with lower firing rates.}
\label{fig:pulse-response-ctrnn}
\end{figure}

\noindent This set of parameters (gain functions and \textbf{Table \ref{Tab:true-values-and-statistics}}) allows the network to have a unique equilibrium state for each stationary input. 
To demonstrate the excitatory and inhibitory characteristics of our model, we can stimulate the model with a square wave (pulse) stimulus as shown in \textbf{Figure \ref{fig:pulse-response-ctrnn}A}. The resultant excitatory and inhibitory neural membrane potential responses ($V_e\left(t\right)$ and $V_i\left(t\right)$) are shown in \textbf{Figure \ref{fig:pulse-response-ctrnn}B} and \textbf{Figure \ref{fig:pulse-response-ctrnn}C}. It can be said that, the network has shown both transient and sustained responses. In \textbf{Figure \ref{fig:pulse-response-ctrnn}D}, the excitatory firing rate response $r_e\left(t\right)$ which is related to excitatory potential as $r_e\left(t\right)=g_e\left(V_e\left(t\right)\right)$ is shown. The response $V_i\left(t\right)$ is slightly delayed which leads to the depolarization of excitatory unit until $t=250$ms. This delay is also responsible from the subsequent re-polarization and plateau formation in the membrane potential of excitatory neuron. The firing rate $r_e\left(t\right)$ is higher during excitation and lower in subsequent plateau and repolarization phases (\textbf{Figure \ref{fig:pulse-response-ctrnn}D}).

\noindent The optimisation of the stimuli requires that the maximum power level in a single stimulus is bounded. This is a precaution to protect the model from potential instabilities due to over-stimulation. In addition, if applied in a real environment the experiment subject will also be protected from such over-stimulations. As the amplitude parameter is assumed positive, assigning an upper bound defined as $A_{max}$ should be enough. This is applied to all stimulus amplitudes (i.e. $A_n$). In this research, a fixed setting of $A_{max}=120\ \mu$A  is chosen. The lower bound is obviously $A_{min}=0$. For the phase $\phi_n$, no lower or upper bounds are necessary as the cosine function itself has already been bounded as $\left(-1\leq\cos\leq1\right)$. The frequencies of the stimulus components are $k^th$ harmonics of a base frequency $f_{base}$ $\left(\omega_k=2\pi{k}\times{f_{base}}\right)$. Since we have a simulation time of $T_{opt}=3$ seconds, we have a reasonable choice of $3.33$ Hz which is in fact equal to $\frac{10}{3}$. So we have chosen an integer relationship between the stimulation frequency and simulation time. The number of stimulus components $N$ is chosen as $N=5$ which is found to be reasonable concerning speed and performance balance. \textcolor{red}{These are reasonable choices specifically for this research because utmost importance is first given to computational performance. The first few simulations showed that choosing $ N $ at a higher setting then $ N=5 $ leads to longer simulation durations which are not desirable. In addition there seems no significant advantage of a larger setting for $ N $. For the base frequency $ f_{base} $, few different values other than $ f_{base}=3.33 $ Hz are tried ($ f_{base}=1 $ Hz, $ f_{base}=5 $ Hz, $ f_{base}=10 $ Hz) but the optimization worked best at the mentioned frequency.} 

It is well known that optimization algorithms such as \emph{fmincon} requires an initial guess of the optimum solution. A suitable choice for the initial guesses can be their assignment from a set of initial conditions generated randomly between the optimization bounds. In the optimization of stimuli, the initial amplitudes can be uniformly distributed between $\left[0,A_{max}\right]$ and phases can be uniformly distributed between $\left[-\pi,\pi\right]$. Although we do not have any constraints on the phase parameter, we limit the initial phase values to a safe assumed range. \textcolor{red}{In the analysis of the optimal phase results we will wrap the resultant phase values to a range $\left[-\pi,\pi\right]$ using \emph{modulo} function. Thus assuming an initial range in the same interval will be meaningful.} 

\noindent We follow a similar strategy for the parameter estimation based on maximum likelihood method. The multiple initial guesses will be chosen from a set of values uniformly distributed between the lower and upper bounds defined in \textbf{Table \ref{Tab:true-values-and-statistics}}.

In \textbf{Section \ref{Sub:jmle-theory}}, one recall from \eqref{eq:joint-likelihood-product} that the likelihood estimation should produce better results when the number of samples (i.e. $M$ in \eqref{eq:joint-likelihood-product}) increases. Because of this fact, the likelihood function is based on data having all spikes generated since the beginning of the simulation. The number of repeats determines $M$. If simulation is repeated $N_{itr}$ times ($N_{itr}$ iterations), one will have an $M$ value of $M=8N_{itr}$ due to the fact that each iteration has $8$ optimal designs sub-steps (with respect to each parameter $\theta_k$. Read \textbf{Section \ref{Sub:oed-theory}}). So, if one has 15 iterations $N_{itr}=15$, $M=120$ which means likelihood has $120$ samples. This also means that optimal design and subsequent parameter estimation will also be repeated $120$ times.

\subsection{Statistics of optimally designed stimuli} \label{sub:statistics-numerical-discuss}

\begin{figure}[H]
\centering
\includegraphics[scale=0.7]{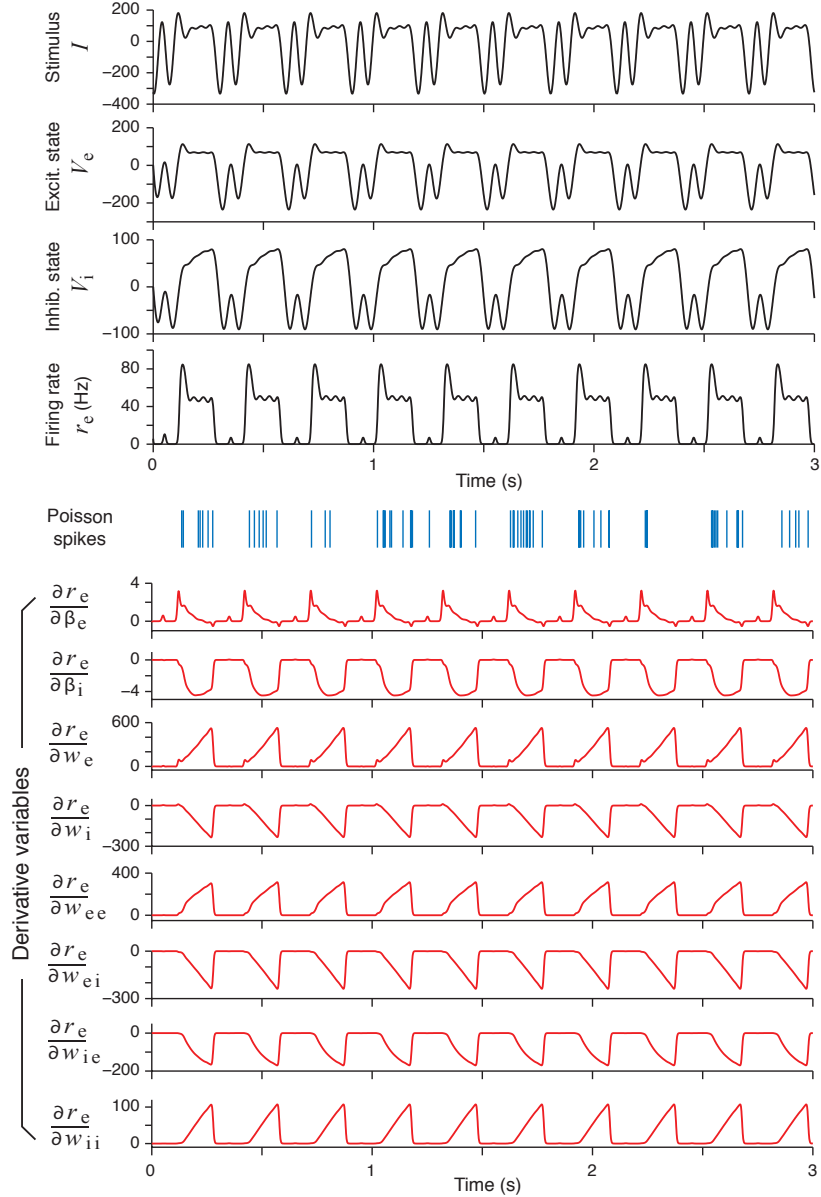}
\caption{An example of an optimally designed stimulus with a duration of  3 sec (top panel), . 
The responses of the excitatory and the inhibitory units in the network are shown below, followed by an example of spike trains generated by an inhomogeneous Poisson process according to the continuous firing rate of the excitatory unit $\left(r_e\right)$. 
Driven by the same stimulus, the response of eight derivatives variables, namely, the derivatives of the firing rate $r_e$ with respect to all the network parameters, are shown as red curves.
These derivatives were solved directly from differential equations \eqref{eq:gradient-evolution-wrt-stimulus-param}. \textcolor{red}{The sensitivity derivatives shown in this figure above are evaluated at the true values of parameters shown in \textbf{Table \ref{Tab:true-values-and-statistics}}. This evaluation is performed to show the level differences between the sensitivity of response to each individual parameter.}}
\label{fig:stimExampleDerivatives}
\end{figure}

Having all necessary information from \textbf{Section \ref{sub:example-details}}, one can perform an optimal design and obtain a sample optimal stimulus and associated neural responses \textbf{Figure \ref{fig:stimExampleDerivatives}}. It is noted that the optimal stimulus in \textbf{Figure \ref{fig:stimExampleDerivatives}} top panel has a periodic variation as it is modelled as a phased cosines form (as it is equivalent to real valued Fourier series).

In addition to the fundamental responses, the second half of \textbf{Figure \ref{fig:stimExampleDerivatives}} displays the variation of the parametric sensitivity derivatives $\frac{\partial{r_e}}{\partial{\theta_i}}$ which are generated by integrating \eqref{eq:core-components-compact} and then substituting to \eqref{eq:derivat-firing-rate-potential-theta}.  
The variation of the sensitivity derivatives support the idea of optimization of the Fisher Information Metric with respect to a single parameter (see \eqref{eq:fisher-wrto-single-param}) \textcolor{red}{as the sensitivity (or gradient) of the Fisher Information Metric in \eqref{eq:fisher-wrto-single-param} with respect to a single parameter $ \theta_k $ varies widely from parameter to parameter. }. It appears that, the weight parameters have a very high sensitivity which are more than $10$ times that of the $\beta_*$ parameters. Also some of the parameters affecting the behaviour of inhibitory (I) unit $\beta_i,w_i$ and also the E-I interconnection weights $w_{ei},w_{ie}$ have a reverse behaviour. This fact appears as a negative values variation of the sensitivity derivatives. Self inhibition coefficient $w_{ii}$ does not show this behaviour as it represent an inhibitory effect on the inhibitory neuron potential (which favours excitation).      

\begin{figure}[H]
\centering
\includegraphics[scale=0.7]{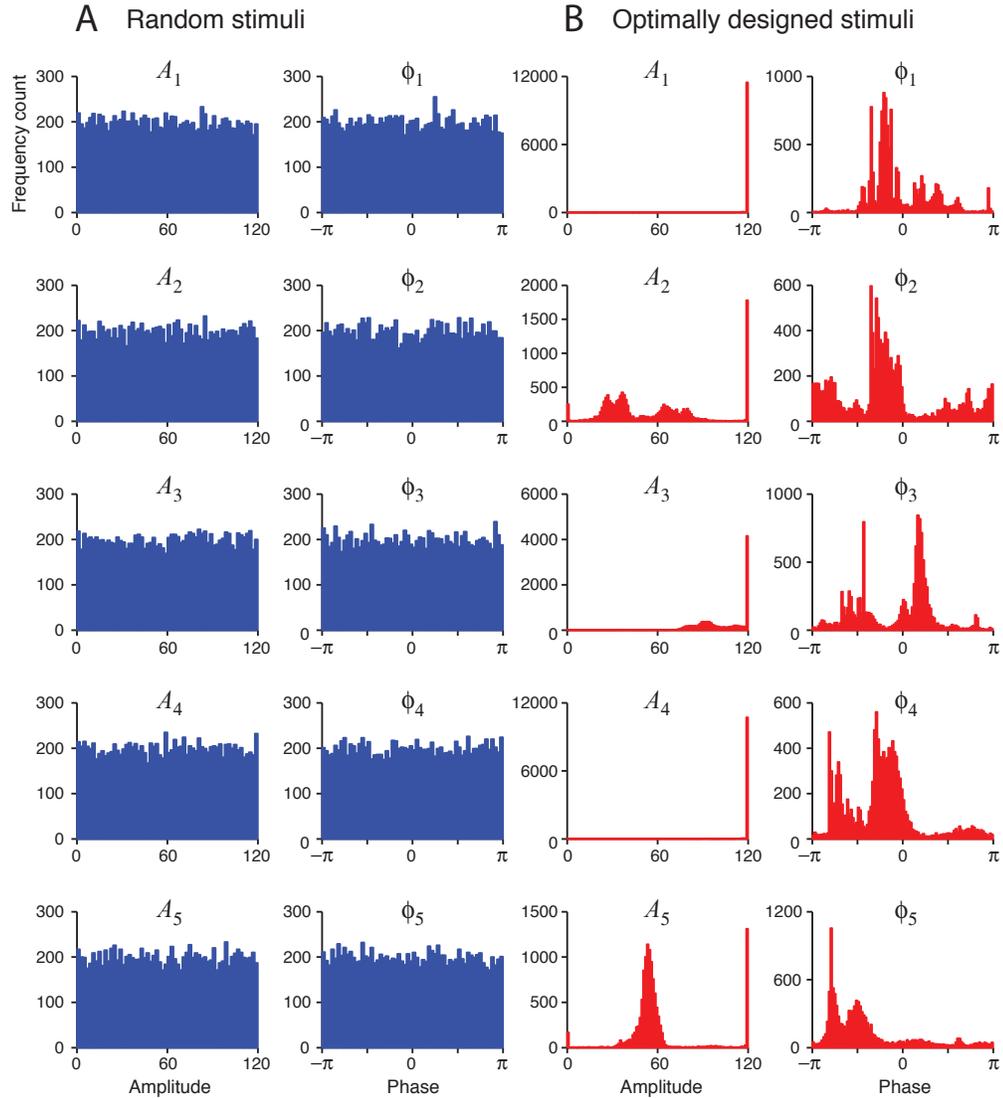}
\caption{Histograms of the stimulus Fourier amplitudes $A_1, \ldots, A_5$ and phases $\phi_1, \ldots \phi_5$. 
(\textbf{A}) 
Random stimuli were generated by choosing their Fourier amplitudes and phases randomly from uniform distributions. 
A total of 12,000 random stimuli were used in each plot.
(\textbf{B})  
The optimally designed stimuli showed some structures in the distributions of their Fourier amplitudes and phases, 
which differ radically from a uniform distribution. A total of 12,000 optimally designed stimuli were used in each plot.}
\label{fig:stimulus-amplitude-histogram}
\end{figure}

One of the most distinguishing result related to optimal design is the statistics of the optimal stimuli (i.e. the amplitudes $A_n$ and phases $\phi_n$). This analysis can be performed by generating the histograms of $A_n$ and $\phi_n$ from available data as shown in \textbf{Figure \ref{fig:stimulus-amplitude-histogram}}. The \textbf{Figure \ref{fig:stimulus-amplitude-histogram}A} shows the flat uniform distribution of the random stimuli amplitudes and phases. This is expected as the stimuli is generated directly from a uniform distribution. Concerning the optimal stimuli, the histograms shown in \textbf{Figure \ref{fig:stimulus-amplitude-histogram}B} reveals that optimal design has a tendency to maximize the amplitude of the stimuli towards the upper bound. This reassures that optimal design tends to maximize the stimulus power which is expected increase the efficiency of the parameter estimation process and it distinguishes optimal stimuli from their randomly generated counterparts. 

\textcolor{red}{
	The results mentioned above coincide the findings of \cite{dimattina2008optimal}. Here, what happens is related to the topological boundary property \cite{mendelson1990introduction} associated with the optimal design results. It is found in \cite{dimattina2008optimal} that, maximum firing rate response is always lies on the topological boundary of the collection of all allowable stimuli provided that the neurons have increasing gain functions and convergent synaptic connections between layers. Another interpretation of this result can be expressed as follows: The maximum and minimum responses of each individual neuron should arise from the stimulus boundary (from all available sets of stimuli) and the entire boundary of the pattern of responses elicited in a particular layer should also a result of the stimulation in the boundary level. This result is valid regardless of the neuron being feed-forward or recurrent. For recurrent neural networks we assume that the network is stable. Some mathematical proofs related to the stated results are also presented in \cite{dimattina2008optimal}. In \textbf{Figure \ref{fig:stimulus-amplitude-histogram}B}, the amplitudes of the Fourier stimulus components are collected at the boundary level which is $ A_{\max}=120 $. This is the topological boundary of the set of stimuli defined by \eqref{eq:cosine-stimulus}.            	
}

\indent Given a dataset consisting of stimulus-response pairs, we can always use maximum-likelihood estimation to fit a model to the data to recover the parameters.
Maximum-likelihood estimation is known to be asymptotically efficient in the limit of large data size, in the sense that the estimation is asymptotically unbiased (i..e,  average of the estimates approaches the true value) and has  minimal variance (i.e., the variance of the estimates approaches the Cram\'er-Rao lower bound).

\begin{figure}
\centering
\includegraphics[scale=0.64]{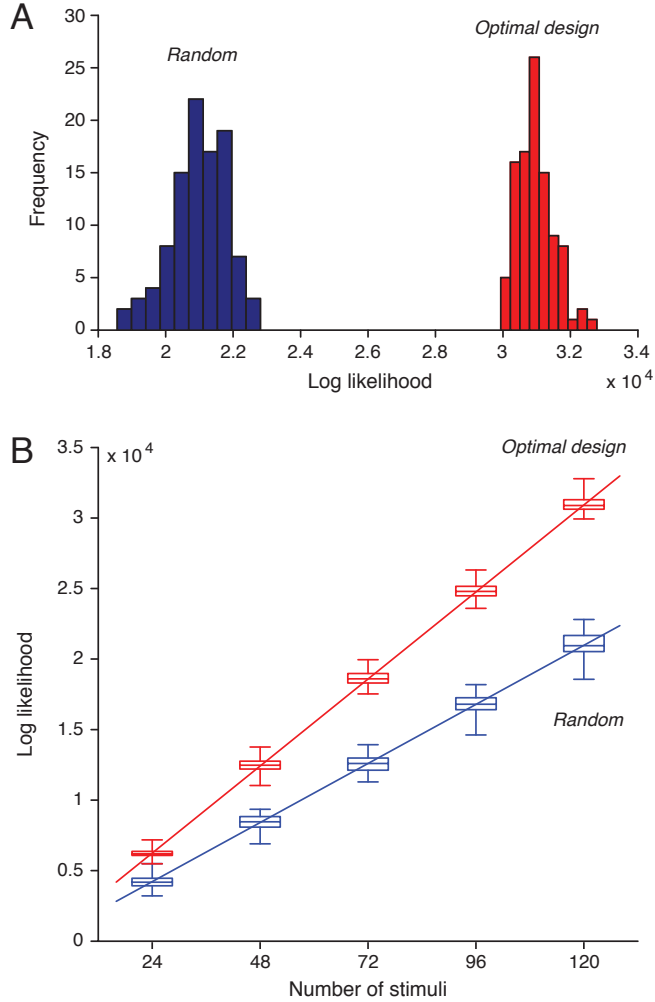}
\caption{Optimally designed stimuli yield greater likelihood values for maximum-likelihood parameter estimation compared to random stimuli.
(\textbf{A})
Histogram of the optimized likelihood values for 100 trials with random stimuli is compared with that for 100 trails with optimally designed stimuli. Note that the best value from the random trials was not even close to the worst value from the optimal design trials.
Here each trial contained a sequence of 120 stimuli, which were generated either randomly or by optimal design.
Each likelihood value was obtained by maximizing the likelihood function in equation xxx using the response data elicited by all 120 stimuli.
(\textbf{B})  
As the number of stimuli increases, the likelihood function also increases, following an approximate linear relationship. 
The optimal design yielded better likelihood values than random stimuli regardless of the number of stimuli. 
The boxplot shows 25\%, 50\% (median) and 75\% percentiles, with the whiskers outside of the boxes indicating the minimum and maximum values. 
The straight lines were obtained by linear regression on the median values.}
\label{fig:likelihood-traces-box}
\end{figure}

\noindent We found that maximum likelihood obtained from the optimally design stimuli was always much better than that obtained from the random stimuli (see \textbf{Figure \ref{fig:likelihood-traces-box}}). It also reveals that, the likelihood value increases as the number of stimuli increases. For any given number of stimuli, the optimally designed stimuli always yielded much greater likelihood value than the random stimuli. The minimum difference between the likelihood values (the minimum from the optimal design and the maximum from the random stimuli based test) was typically about two times greater than the standard deviation of either estimates except for the case with $24$ samples $\left(N_{itr}=3 \& M=24\right)$. Even in this case, this violation appear only on one sample. In addition, it can be easily deduced from the box diagram in \textbf{Figure \ref{fig:likelihood-traces-box}} that the difference between $75$\textsuperscript{th} and $25$\textsuperscript{th} percentiles ($3$\textsuperscript{st} and $1$\textsuperscript{rd} quartiles) yield a value which is larger than three times the standard deviation of either estimate.  The standard deviations of the maximum likelihood values are also larger in the random stimuli based tests. Those results are certain evidences of the superiority of an optimal design over the random stimuli based tests. The difference between the maximum values of the two likelihoods (from optimal and random stimuli) becomes more significant as the number of samples $M$ increases. 

It would also be convenient to stress the fact that the greater the likelihood, the better the fitting of the data to the model. This fact is demonstrated by two regression lines imposed on the box diagram \textbf{Figure \ref{fig:likelihood-traces-box}}. One of those lines correspond to the optimal design and the other correspond to the random stimuli based tests.   
Both regression lines path through the origin point (0, 0) approximately. This means that the ratio of the log likelihood values in the two cases is approximately a constant, regardless of the number of stimuli. The regression lines are represented by equations $l=257.0896M+76.1500$ for optimal and $174.5375M+31.1000$ for the random stimuli. So the two lines have a slope ratio of approximately $1.4730$. The significance of this number can be explained by a simple example. If one desires to attain the same level of likelihood with $M=120$ optimally designed stimuli, the required number of random stimuli to be generated is equal to a value about $180$. 

Another statistical comparison of the maximized likelihoods can be performed by Wilcoxon rank-sum tests \citep{mann1947test,gibbons2011nonparametric,hollander2013nonparametric}. When applied, one will be able to see the difference which is highly significant. Regardless of the number of samples the p-values remained at least $10^{-30}$ times smaller than the widely accepted probability significance threshold of $p=0.05$ or $5\%$.     

\begin{figure}[H]
\centering
\includegraphics[scale=0.7]{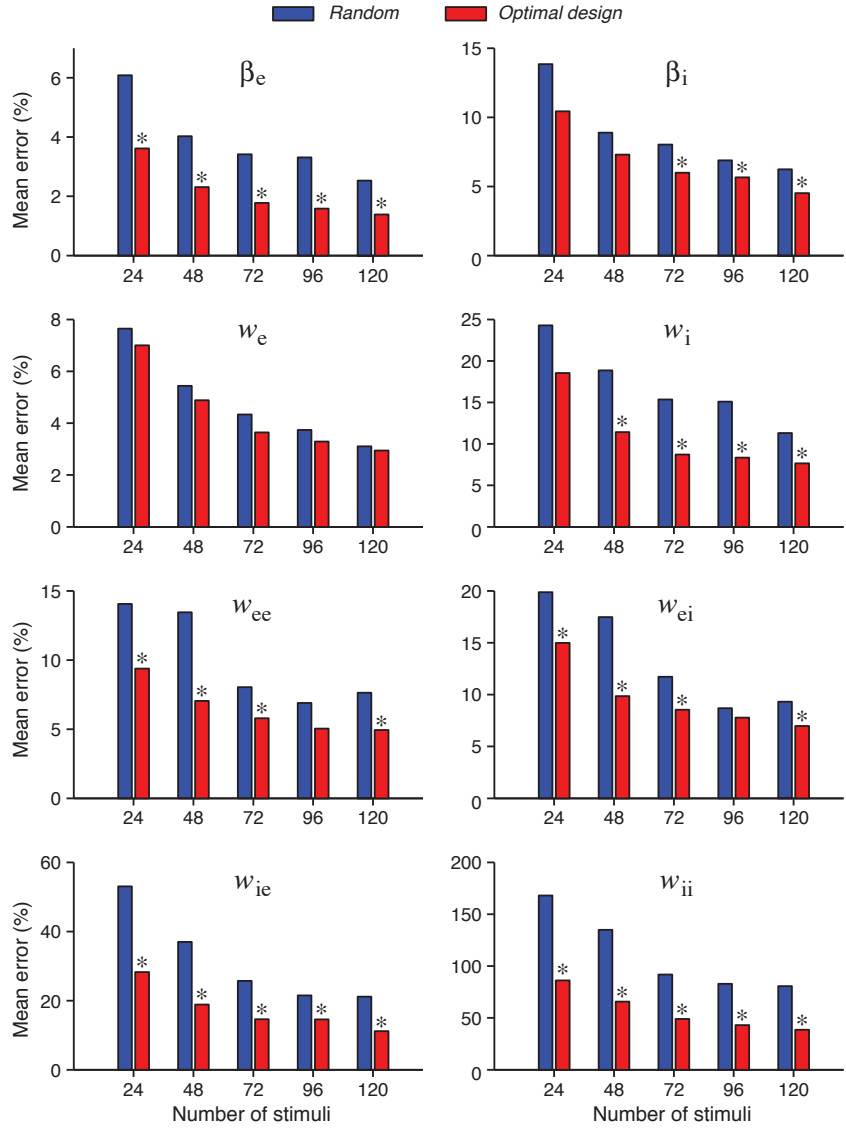}
\caption{Errors of individual network parameters obtained by maximum likelihood estimation tended to decrease as the number of stimuli increased. The optimal design yielded smaller errors for all parameters in all the cases tested, although the differences were not always statistically significant.An asterisk ($*$) at the top of an optimal design bar indicates that the difference from the neighboring random bar is statistically significant at the level $p<0.05$ in the ranksum test. }
\label{fig:mean-error-bars}
\end{figure}

The likelihood function provides an overall measure of how well a model fits the data. We have also tested the mean errors of individual parameters relative to their true values. The main finding is that, for each individual parameter, the error is typically smaller for the optimally designed stimuli than the error for the random stimuli. This result can be observed from the bar charts presented in \textbf{Figure \ref{fig:mean-error-bars}}. The heights of the bars show the mean error levels for randomly and optimally generated stimuli respectively. One can get the benefits of the rank-sum test on the statistical properties of the parameter estimates. Computation of rank-sum p-values for each individual parameter corresponding to the case of $120$ samples $\left(M=120\right)$ yields:

\begin{equation}
\begin{array}{c}
p\left(\beta_e\right)=4.4711\times10^{-5}\\
p\left(\beta_i\right)=9.0988\times10^{-3}\\
p\left(w_e\right)=9.8928\times10^{-1}\\
p\left(w_i\right)=6.8591\times10^{-3}\\\
p\left(w_{ee}\right)=1.9874\times10^{-3}\\
p\left(w_{ei}\right)=5.5709\times10^{-3}\\
p\left(w_{ie}\right)=2.2302\times10^{-5}\\
p\left(w_{ii}\right)=2.8541\times10^{-8}
\end{array}
\end{equation}  

\noindent The above result showed that for $M=120$ the differences are statistically significant for $7$ parameters. For different values one can refer to \textbf{Figure \ref{fig:mean-error-bars}}. In this illustration, the statistical significance of the difference between optimal and random stimuli based tests are indicated as an asterisk placed above the bars. Any of the cases with asterisk means that, the associated sample size led to a result where optimal design is significantly better.

\noindent Here an interesting result is the statistical non-significance of the third parameter $w_e$ regardless of the sample size $M$. This might be associated with the parameter confounding phenomenon that is discussed in \textbf{Section \ref{sub:parameter-confo}}.  

\noindent For convenience, one can see the mean values and standard deviations of the estimates obtained from $100$ optimal and random stimuli in \textbf{Table \ref{Tab:true-values-and-statistics}}. These are obtained with $M=120$ samples. In general, mean values seem to be comparable for both stimuli however the standard deviations of estimates from optimal stimuli are smaller then that obtained from random stimuli. The mean values of parameters $w_{ie}$ and $w_{ii}$ are also closer to the true values when obtained from optimal stimuli.
\subsection{Problem of local maxima during optimization}\label{sub:local-maxima-discussion}

We need optimization in two places: maximum likelihood estimation of parameters, and optimal design of stimuli. Due to speed and computational complexity considerations one needs to utilize gradient base local optimizers such as MATLAB\textsuperscript{\circledR} \emph{fmincon}. These algorithms often needs initial guesses and not all of the initial guesses will converge to a true value. This issue may especially appear in the cases where the objective function involves dynamical model (differential or difference equations). In such cases, the problem of multiple local maxima might occur in the objective function which often requires multiple initial guesses to be provided to the solver. So an analysis on this issue may reveal useful information.      

Suppose there are $n$ repeats or starts from different initial values. Let $p$ be the probability of finding the ``correct'' solution in an individual run with random initial guess. Then in $n$ repeated runs, the probably that at least one run will lead to the ``correct'' solution is

\begin{equation}
	\mbox{Prob (``correct'')}=1-(1-p)^n
\end{equation}

\noindent The probability $p$ can be estimated from a pairwise test or directly from the values of the likelihood and Fisher Information Metric. In the test of the likelihood function, one can achieve the goal by starting the optimization from $K$ different initial guesses and checking the number of solutions which stay in an error bound $10\%$ for each individual parameter with respect to the solution leading to the highest likelihood value. In other words, to pass the test the following criterion should be satisfied for each individual parameter $\theta=\theta_i$ in \eqref{eq:theta-ctrnn-param}:

\begin{equation}
\frac{\left|\hat{\theta}_{best}-\hat{\theta}\right|}{\theta}\leq{10}{\%} 
\label{eq:optimal-error-crit}
\end{equation} 

\noindent where $\hat{\theta}_{best}$ is the local optimum solution having the highest objective (likelihood) value and $\hat\theta$ is the estimated value of $\theta$. If the above is satisfied for all $\theta_i$, this result is counted as one pass.   

\noindent So for maximum likelihood estimation with $M=120$, the data suggests a probability of $p\approx 0.85$ and to get a $99\%$ correct rate we will only need $n=3$ repeats. This is a result obtained from $10$ multiple initial guesses per $20$ different stimuli configurations (total of $200$ occurrences). Here, we have a high probability of obtaining a global maxima and thus we may get rid of multiple initial guesses requirement in the estimation of $\theta$. 

For the optimal design part the problem is expected to be harder as the stimulus amplitudes tend to the upper bound $A_{max}$. \textcolor{red}{This should be due to the increasing level of response as the stimulus approaches the upper bound}. It should also be remembered that, the Fisher Information measure is computed with respect to a single parameter $\theta_i$ as shown in \eqref{eq:fisher-wrto-single-param}. Thus it will here be convenient to analyse the respective Fisher Information metric $U_k\left(\bf{x},\theta_i\right)$ as defined w.r.to each parameter $\theta_i$. Like in the case of likelihood analysis, we do the analysis on $20$ samples (i.e. $20$ stimuli) separately for amplitudes $\left(A_n\right)$ and phases $\left(\phi_n\right)$. The criterion for passing the test is similar to that of \eqref{eq:optimal-error-crit} after replacing $\hat\theta$ by $\bf x$ from \eqref{eq:x} and $\hat{\theta}_{best}$ by $\mathbf{x}_{best}$ with $\mathbf{x}_{best}$ being the stimulus parameter yielding the largest value of $U_k\left(\bf{x},\theta_i\right)$. Note that, since we do not have any concept of "true stimulus parameters" we will use $\mathbf{x}_{best}$ in the denominator of \eqref{eq:optimal-error-crit}. 

\noindent After doing the analysis for amplitudes $A_n$, one can see that highest probability value $p=0.8625$ is obtained for $U_1$ whereas the smallest value is obtained for $U_5$ as $p=0.175$. The second and third smallest values are $U_3$ and $U_6$ having $p=275$ and $p=0.35$ respectively. The indices $1$,$3$,$5$ and $6$ correspond to the $\beta_e$,$w_e$,$w_{ee}$ and $w_{ei}$. This means that the lowest probability values occur at the parameters $w_e$,$w_{ee}$ and $w_{ei}$. This result might be interesting as those three parameters have strong correlations at least with one other network parameter (see \textbf{Section \ref{sub:parameter-confo}}). The required number of repeats appears to be $n=23$ for the worst case ($p=0.175$ for $U_5$).               

\noindent For the phase parameters $\phi_n$, different initial conditions lead to different values. This is an expected situation \textcolor{red}{as \emph{fmincon} or similar functions are local optimizers. In fact, the stochastic nature of the global optimizing routines such as the genetic algorithms or simulated annealing will also lead to a similar result when initial populations are provided after each run.} Because of this outcome, the solution yielding the largest value of Fisher Information Metric $\left(U_k\right)$ among all runs with different initial conditions should be preferred in the actual application.   

\noindent Modern parallel computing facilities will ease the implementation of optimization with multiple random guesses.

\subsection{Parameter confounding}\label{sub:parameter-confo}

\begin{figure}[H]
\centering
\includegraphics[scale=0.62]{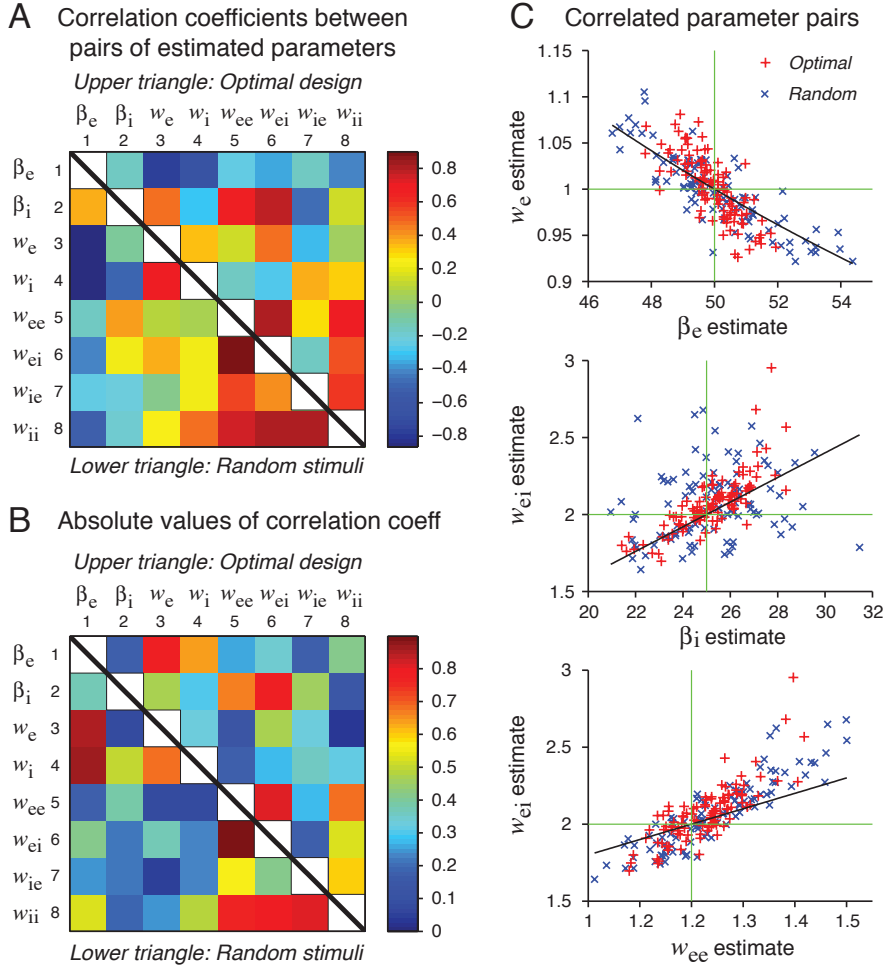}
\caption{Some network parameters are approximately confounded. Here confounding means that a change of one parameter can be compensated by a proper change of another parameter, such that the stimulus-response relation of the whole network is kept approximately the same. 
(\textbf{A})
Correlation coefficient matrix of all possible pairs of network parameters obtained by maximum likelihood estimation from either optimal design data or random stimulus data. Since each matrix is symmetric, only a half needs to be shown. Here data from optimal design trials are shown in the upper triangle, whereas data from the random trials are shown in the lower triangle. Each data point was based on 100 repeated trials each containing a sequence of 120 stimuli.
(\textbf{B})
Same data as in panel A, except that the absolute values are shown.
(\textbf{C})
The three pairs of parameters with the highest correlation coefficients in the upper triangle in panel B  are shown in the scatter plots. Data from both optimal design and random trials are shown. Black curves are theoretical predictions according to equations \eqref{eq:confound13}, \eqref{eq:confound26} and \eqref{eq:confound56}. Green crosshairs are centered at the true parameter values. }
\label{fig:confounding-plots}
\end{figure}

\indent The errors of some parameters tend to be correlated (see \textbf{Figure \ref{fig:confounding-plots}}).   

Parameter confounding may explain some of the correlations. The idea is that different parameter may compensate each other such that the network behaves in similar ways, even though the parameter values are different. It is known that in individual neurons, different ion channels may be regulated such that diverse configurations may lead to neurons with similar neuronal behaviours in their electrical activation patterns \citep{prinz2004similar}. Similar kind of effect also exists at the network level \citep{dimattina2010modify}.

Here we will consider the original dynamical equations and demonstrate how parameter confounding might arise. We first emphasize that different parameters in our model are distinct and there is no strict confounding at all. The confounding is approximate in nature.

\noindent From the correlation analysis on the optimal design data (\textbf{Figures \ref{fig:confounding-plots}A} and \textbf{\ref{fig:confounding-plots}B}), three pairs of parameters stand out with the strongest correlations. 
These pairs are $(\beta_e$, $w_e)$, $(\beta_i, w_{ei})$ and $(w_{ee},w_{ei})$.

\noindent We will offer an intuitive heuristic explanation based on the idea of parameter confounding.

\noindent We here rewrite the dynamical equations for convenience as shown below: 
\begin{eqnarray}
	\label{eq:Ved}
	\dot{V_e} &=& -\beta_e V_e+\beta_e\left\{w_{ee}g_e(V_e)-w_{ei}g_i(V_i)\right\}+\beta_e w_e I\\
	\dot{V_i} &=& -\beta_i V_i+\beta_i\left\{w_{ie}g_e(V_e)-w_{ii}g_i(V_i)\right\}+\beta_i w_i I
	\label{eq:Vid}
\end{eqnarray}

\subsubsection*{{Example 1:} Confounding of the parameter pair $(\beta_e$, $w_e)$.}

The external stimulus $I$ drives the first equation \eqref{eq:Ved} through the weight $\beta_e w_e$. If this product is the same, the drive would be the same, even though the individual parameters are different. For example, if $\beta_e$ is increased by 10\% from its true value while $w_e$ is decreased by 10\% from its true value, then the product stays the same, so that the external input provides the same drive to \eqref{eq:Ved}. Of course, any deviation from the true parameter values also leads to other differences elsewhere in the system.
Therefore, the confounding relation is only approximate and not strict. 
This heuristic argument gives an empirical formula: 
\begin{equation}
	\hat{\beta}_e \hat{w}_e=\beta_e w_e \label{eq:confound13}
\end{equation}
where $\beta_e$ and  $w_e$ refer to the true values of these parameters, whereas  $\hat{\beta}_e$ and $\hat{w}_e$ refer to the estimated values.

\subsubsection*{{Example 2:} Confounding of the parameter pair $(w_{ei}, \beta_i$).}

These two parameters appear separately in different equations, namely, $w_{ei}$ appearing only in \eqref{eq:Ved} while $\beta_i$ appearing only in \eqref{eq:Vid}. 
To combine them, we need to consider the interaction of these two equations. 
To simplify the problem, we consider a linearised system around the equilibrium state:
\begin{eqnarray}
	\label{eq:Vedlin}
	\dot{V_e} &=& -\beta_e V_e+\beta_e\left\{w_{ee}k_e V_e-w_{ei}k_i V_i\right\}+\beta_e w_e I+C_e\\
	\dot{V_i} &=& -\beta_i V_i+\beta_i\left\{w_{ie}k_e V_e -w_{ii}k_i V_i\right\}+\beta_i w_i I+C_i
	\label{eq:Vidlin}
\end{eqnarray}
where $k_e$ and $k_i$ are the slopes of the gain functions, and $C_e$ and $C_i$ are extra terms that depend on the equilibrium state and other parameters.
Note that $V_i$ appears in \eqref{eq:Vedlin} only once, in the second term in the curly brackets. Since $V_i$ also satisfies 
\eqref{eq:Vidlin},  we solve for $V_i$ in terms of $\dot V_i$ from \eqref{eq:Vidlin} and find a solution of the form: $V_i=c \dot V_i/\beta_i +a$ where $c$ is a constant.
Substitution into \eqref{eq:Vedlin} shows that the parameter combination $w_{ei}/\beta_i$  scales how strongly $\dot V_i$  influences this equation.
Thus we have a heuristic confounding relation:
\begin{equation}
	\hat w_{ei}/\hat\beta_i=w_{ei}/\beta_i \label{eq:confound26}
\end{equation}

\subsubsection*{{Example 3:} Confounding of the parameter pair $(w_{ee},w_{ei})$.}

These two parameters both appear in the curly brackets in \eqref{eq:Ved}. We have a heuristic confounding relation: 
\begin{equation}
	\hat w_{ee}g_e(\bar V_e)-\hat w_{ei}g_i(\bar V_i)=w_{ee}g_e(\bar V_e)-w_{ei}g_i(\bar V_i)
	\label{eq:confound56}
\end{equation}
where $\bar V_e$ and $\bar V_i$ are the equilibrium states.
If this equation is satisfied, we expect that 
the term in the curly brackets in \eqref{eq:Ved} would be close to a constant (the right-hand side of \eqref{eq:confound56}) whenever the state $V_e$ and $V_i$ are close to the equilibrium values. When the state variables vary freely, we expect this relation to hold only as a very crude approximation.

Simulation results show that these three confounding relation can qualitatively account for the data scattering. 
The random data follow the same pattern (see \textbf{Figure \ref{fig:confounding-plots}}C) although they appear to have more scattering compared to the optimal design based data. 
Although the confounding relations are not strictly valid, their offer useful approximate explanations that are based on intuitive argument and are supported by the data as shown in \textbf{Figure \ref{fig:confounding-plots}}.

The theoretical slopes are always smaller, suggesting that the heuristic theory only accounts for a portion of the correlation. 
It is likely that there are approximate confounding among more than those three pairs of parameters namely $(\beta_e$, $w_e)$, $(\beta_i, w_{ei})$ and $(w_{ee},w_{ei})$.

\section{Discussion} \label{sub:discussion}
\subsection{Summary of the results}

The main results of this paper can be summarized as follows. 

\begin{enumerate}
\item We have implemented an optimal design algorithm for generating stimuli that can efficiently probe the E-I network, which describes the dynamic interaction between an excitatory neuronal population and an inhibitory neuronal population. The data has been used to model the auditory system etc.
\item The dynamical network allows both transient and sustained response components (\textbf{Figure \ref{fig:pulse-response-ctrnn}})
\item Derivatives are computed directly by differential equations derived from the original system (\textbf{Figure \ref{fig:stimExampleDerivatives}})  
\item Optimally designed stimuli have patterns. The amplitude tend to be saturated, which can lead to faster search because only boundary values need to be checked (\textbf{Figure \ref{fig:stimulus-amplitude-histogram}})
\item The optimally designed stimuli elicited responses that have more dynamic range (more variance in the histogram of amplitude of the stimulus waveform $I\left(t\right)$)
\item The optimal design yield much better parameter estimation in term of likelihood (\textbf{Figure \ref{fig:likelihood-traces-box}}), and in the errors of the individual parameters (\textbf{Figure \ref{fig:mean-error-bars}}).
\item We studied parameter confounding \textcolor{red}{which is thought to be a major source of the error} (\textbf{Figure \ref{fig:confounding-plots}}).
\item Significance of the results can readily lead to practical applications. For example in the modelling of the auditory networks known works all used stationary sound stimuli. However, it is beneficial to include time dependency as realistic neurons and their networks have several dynamic features. 
\end{enumerate}

\subsection{Defence of this Research}

\subsubsection*{Choice of the model:}
The chosen model is an appropriate and simple model which has only two neurons with one being excitatory and one being inhibitory. This choice is reasonable in the context of this research as the availability of previous work targeting a similar goal is quite limited. In addition, the known ones mostly based on the static feed-forward network designs. In order to verify our optimal design strategy we have to start from a simpler model. The theoretical development and results of this work can very easily be adapted to alternative and/or more complicated models. Here the most critical parameter is the number of data generating neurons (neurons where the spiking events are recorded by electrode implantations) and the computational complexity. The former is a procedural aspect of the experiment whereas the latter is directly related to the instrumentation. In addition, the universal approximating nature of the CTRNN's is an advantage on this manner.  

\subsubsection*{Speed of computation:}\label{sub:computational-speed-issues}
In this work, we aim at investigation of the computational principles without trying the maximizing the computation speed. 
Right now on a single PC, having a Intel\textsuperscript{\circledR} Core\texttrademark i7 processor with $6$ cores, the  computational duration for optimization of a single stimulus is about $15$ minutes. Surely, this is an average value as the number of steps required to converge to an optimum depends on certain conditions such as the value of objective value, gradient, constraint violation and step size. A similar situation exist for the optimization of the likelihood. However in this case, the optimization times of the likelihood will vary due to its increasing size as all the historical spiking is taken into account (see \eqref{eq:complete-likelihood-compact}) leading to a function with gradually increasing complexity. Although that is not the only fact contributing to the computational times, the average duration of optimization tend to increase with the size of likelihood $M$. An average value for the observed duration of likelihood optimization is $38$ minutes. As a result optimization of one stimulus and subsequent likelihood estimation requires a duration of about $53$ minutes. This is approximately one hour. So one complete run with a sample size of $M=120$ is completed in a duration about $28$ hours. Changing the value of the sample size $M$ will have a direct influence on the system computation time. For example, the duration of one complete run will reduce to a value about $12$ hours when $M=24$. This is an expected situation as there will be a reduced number of the summation terms in the likelihood function. The reduction in the duration of the optimal design algorithm is only based on the reduction in the number of trials which is just equal to the sample size $M$. So based on these findings, one will need a speed-up in order to adapt this work to an experiment. There are several ways to speed up the computation:
\begin{enumerate}
\item Using a large time step: In this work we have integrated the equations using a time step of $1$-millisecond or $0.001$-seconds. This value may be increased to levels as high as $0.01$-seconds. This modification will have a little contribution to the speed of computation. The benefits will most likely be from the optimal design part due to the manipulation of a single interval (no consideration of past/historical spike trains). However, the higher the time step the lower the accuracy of the estimates and the optimality of the stimuli. This main contributor to this fact is the spike generation algorithm where the accuracy of the locations of some spikes are lost when a coarse integration interval is applied.
\item Using and/or developing streamlined optimization algorithms: This can be a subject of a new project on the same field. This development is expected to have a considerable contribution to the computation time without any trade-offs over performance.
\item Generating the stimuli as a block rather than one by one: This is also a potential topic for a new project. This is expected to reduce the optimal design time without losing performance.
\item Employment of larger cluster computing systems (or high performance computing systems (HPC)) having more than $100$ CPU cores: Though the most sophisticated and expensive solution, it is the best approach to cure the overall computational burdens and transform the theoretical only study to an experiment adaptable one. 
\item Porting the algorithms to a lower level programming language such as C/C++ or FORTRAN may help in speeding up the computation. If an efficient and stable numerical differentiation algorithm can be employed in this set-up, another optimality criterion such as \emph{D} or \emph{E} optimality can be used in the computation of the Fisher Information Metric which might help in reducing the number of steps (i.e $M$ and/or $N_{itr}$ values).    
\item Knowing the fact that the optimal stimulus amplitudes $A_n$ tend to the upper boundary $A_{max}$ (remember from \textbf{Figure \ref{fig:stimulus-amplitude-histogram}B}). All the amplitudes can be set to same value as $A_n=A$ and only $A$ is computed from optimization. This setting can be helpful for speeding up the optimization time during an actual experiment. However, it should be verified by simulation whether this choice is meaningful for an actual experiment. In this case, one may not need many repeats as only the statistics of the stimuli is required. This occurrence is quite common in the simulation results thus we do not expect a performance degradation when this change is applied to the stimuli characterization.      
\end{enumerate}

\noindent With the above adaptations, it is expected that we are within reach to reduce the computation time for  each 3-second stimuli to less than 3 seconds. In addition one can has the following options which are related to tuning of the algorithms used in this research:
\begin{enumerate}
\item The parameter related to the sample size $(M)$ might be reduced. That is an examined situation in this study. Optimal design seems to yield a better estimation performance with a reduced $M$ compared to random stimuli case with same $M$. However, the former will lead to a slightly increased computational time. However, the trade-off is not very direct. For example, a random stimuli based simulation with $M=120$ samples requires a longer run than a simulation based on an optimally designed stimulus with $24$ samples. This is fairly a good trade-off.
\item Another option to increase the computational performance might be the reduction of the cut-off points in the optimization algorithm such as the first order optimality measure (tolerance of the gradient) and the step-size. This will result in a faster computation but this approach may bring out questions on the accurate detection of the local minimums among which the best one is chosen (both in OED and likelihood optimization). For the \texttt{fmincon} algorithm in MATLAB the first order optimality tolerance and step-size might both be shifted from $1 \times 10^{-09}$ to $1 \times 10^{-06}$. This tuning brings improvement in the computational duration about 10\% without a considerable performance loss. However, if one has a HPC supported computational environment it is strongly recommended not to modify these settings. 
\end{enumerate}

\subsubsection*{Future Issues:}

The above network is rather a simpler example to demonstrate the optimal design approach and its computational challenges. However more features can be brought to this research concerning efficiency and applicability to an actual experiment.  

\begin{enumerate}
\item Speed up issues: The tasks related to speed of computation discussed in \textbf{Section \ref{sub:computational-speed-issues}} may be a separate project to be developed on top this research.
\item Large number of neurons may be considered together with multiple stimulus inputs and response data collection from multiple neurons (both excitatory and inhibitory groups of neurons). 
\item More complex stimulus structures may be utilized. This can be achieved by increasing $N$ in \eqref{eq:cosine-stimulus} or considering different stimulus representations other than phase cosines. 
\item In this research, the primary goal was the estimation of the network weights $w_{**}$ and time constants $\tau_*$. However, it will be interesting to test the methodology for its performance in estimation of firing thresholds and slopes. (i.e. the parameters $a_j$ and $h_j$ in \eqref{eq:sigmoid-general})   
\item Some more realistic details like plasticity can be included to obtain a model describing the synaptic adaptation. Although it is expected to be a harder problem, the method takes fewer stimuli and should be faster. 
\item \label{it:optimality} \textcolor{red}{In this research, the optimal design process is performed by maximization of the Fisher Information Metric with respect to a single parameter $ \theta_k $ which are individual elements of the main diagonal of the Fisher Information Matrix. This is at best close to A-Optimality measure of Optimal Design}. However, it is stated in \cite{khinkis2003optimal} that, D-Optimality brings an advantage that the optimization will be immune to the scales of the variables. On the contrary, it is also stated in the same source that this mentioned fact is not true for A- and E-Optimality criteria in general. The sensitivity to scaling of the variables lead to another issue that the confounding of the parameters brings certain problems about the bias and efficiency of the estimates. So it will be quite beneficial to see the results obtained from the same research with the optimal designs performed by D-Optimal and other measures of Fisher Information Metric such as E- and F-Optimality. As these will require a new set of computations it will be better to include them in a future study.
\item Similar to the discussion in \textbf{Article \ref{it:optimality}} above, the methodology of optimization in optimal designs and likelihood optimization should be considered. Current work involves evaluation of gradients for the sake of faster computation. However, this requires larger efforts in the preparation as one should develop a specific algorithm to compute the evolution of the gradients satisfactorily. This is also required for a speed-up. With the availability of a high performance computing system, other optimization methods such as simulated-annealing, genetic algorithms and pattern search might be employed instead of the local minimizers such as \emph{fmincon} of MATLAB. These algorithms may help in searching for a better optimal stimulus.  
\item After a sufficiently fast simulation is obtained an experiment can be performed involving a living experimental subject. The mapping of the actual sound heard by the animal during the course of experiment to the optimally designed stimulus is a critical issue here and will also be a part of the future related research.  
\end{enumerate}







\end{document}